\documentclass[prd, superscriptaddress ,amsmath,amssymb, nofootinbib]{revtex4}

\pdfoutput=1
\usepackage{mathrsfs}
\usepackage {amsmath, amssymb}
\usepackage[usenames,dvipsnames]{color}
\usepackage{hyperref}
\usepackage[pdftex]{graphicx}
\usepackage{subfigure}
\graphicspath{{figs/}}
\usepackage{bbm}
\usepackage{mdwlist} 
\usepackage[normalem]{ulem}
\usepackage{comment}
\usepackage{mathtools}
\usepackage{sidecap}

\advance \topmargin by -\headheight
\advance \topmargin by -\headsep
\evensidemargin \oddsidemargin
\def\simgt{\mathrel{\lower2.5pt\vbox{\lineskip=0pt\baselineskip=0pt
           \hbox{$>$}\hbox{$\sim$}}}}
\def\simlt{\mathrel{\lower2.5pt\vbox{\lineskip=0pt\baselineskip=0pt
           \hbox{$<$}\hbox{$\sim$}}}}

\def\beq{\begin{equation}}
\def\eq{\end{equation}}
\def\eeq{\end{equation}}

\newcommand{\mymatrix}[1]{\begin{pmatrix} #1 \end{pmatrix}}

\definecolor{jeremycolour}{rgb}{0.2,0.5,0.6}
\newcommand{\jm}[1]{\textcolor{jeremycolour}{[jeremy: #1]}}
\newcommand{\mhidden}{m_{\gamma'}}

\begin{document}

\title{
A Parametrically Enhanced Hidden Photon Search
}

\author{Peter W. Graham}
\affiliation{Stanford Institute for Theoretical Physics, Department of Physics, Stanford University, Stanford, CA 94305}

\author{Jeremy Mardon}
\affiliation{Stanford Institute for Theoretical Physics, Department of Physics, Stanford University, Stanford, CA 94305}

\author{Surjeet 
Rajendran}
\affiliation{Stanford Institute for Theoretical Physics, Department of Physics, Stanford University, Stanford, CA 94305}
\affiliation{Berkeley Center for Theoretical Physics, Department of Physics, University of California, Berkeley, CA 94720}

\author{Yue Zhao}
\affiliation{Stanford Institute for Theoretical Physics, Department of Physics, Stanford University, Stanford, CA 94305}

\begin{abstract}
\vspace{5pt}
Many theories beyond the Standard Model contain hidden photons.
A light hidden photon will generically couple to the Standard Model
through a kinetic mixing term, giving a powerful avenue for detection using ``Light-Shining-Through-A-Wall"-type transmission experiments with resonant cavities.
We demonstrate a parametric enhancement of the signal in such experiments, 
resulting from transmission of the longitudinal mode of the hidden photon.
While previous literature has focused on the production and detection of transverse modes, the longitudinal mode allows a significant improvement in experimental sensitivity.
Although optical experiments such as ALPS are unable to take useful advantage of this enhancement, the reach of existing microwave cavity experiments such as CROWS is significantly enhanced beyond their published results.  Future microwave cavity experiments, designed with appropriate geometry to take full advantage of the longitudinal mode, will provide a powerful probe of hidden-photon parameter space extending many orders of magnitude beyond current limits, including significant regions where the hidden photon can be dark matter.
\end{abstract}

\maketitle
\vspace{-25pt}

\tableofcontents

\section{Introduction}

For much of the past century, progress in particle physics has occurred through discoveries at high energies.
This route was made possible through the development of ever more powerful colliders, culminating in the Large Hadron Collider. This route though is limited by our ability to make such machines. However, physics at the highest scales can sometimes reveal itself through low energy manifestations. Probes of such phenomena can  offer a glimpse of nature at scales that cannot be directly reached through traditional, high energy accelerator experiments. These manifestations come with a price - the effects of these phenomena on the standard model are heavily suppressed, and can only be probed through high precision. A well known example of such a phenomenon is the decay of the proton, which can be rendered unstable, albeit over very long time scales, by physics at the unification scale. Ultraviolet physics can also lead to the existence of light particles with masses much less than the typical standard model scales $\sim$ 100 GeV, with highly suppressed couplings.  One example is an axion.

A light, massive vector boson or ``hidden photon" is another well motivated example of such a particle.
Light hidden photons emerge naturally in many scenarios, often associated with light hidden sectors, and have received a great deal of theoretical interest (see e.g.~\cite{Holdom:1986eq, Pospelov:2007mp, Abel:2008ai, ArkaniHamed:2008qn, ArkaniHamed:2008qp, Pospelov:2008zw, Goodsell:2009xc, Arvanitaki:2009hb, Jaeckel:2010ni, Essig:2010ye, Ringwald:2012hr}).
A hidden photon is also an interesting dark matter candidate~\cite{Nelson:2011sf, Arias:2012az}\footnote{See also~\cite{future:HPDM}, which will clarify and extend the results of the previous literature. We note in passing that a light hidden photon is perfectly consistent with a high scale of inflation, such as the value $H_I\!\sim\!10^{14}$~GeV possibly indicated by recent results from BICEP2~\cite{Ade:2014xna} -- see~\cite{future:HPDM} for more details.}.
The most generic way in which a hidden photon interacts with the standard model is through kinetic mixing with the photon~\cite{Holdom:1985ag}. This is particularly natural since kinetic mixing is a dimension 4 operator, and therefore has unsuppressed low energy effects even if it is generated in the far ultraviolet. The Lagrangian that describes this theory is
 \begin{equation}
\mathcal L = -\frac{1}{4} \left( f_{\mu \nu} f^{\mu \nu} + f'_{\mu \nu} f'^{\mu \nu} - 2\varepsilon f_{\mu \nu} f'^{\mu \nu}\right) + \frac{1}{2} m_{\gamma'}^2 a'_\mu a'^\mu - e \, a_\mu j_{EM}^\mu \, ,
\label{KineticMixing}
\end{equation}
where $a_{\mu}$ (along with its field strength $f_{\mu\nu}$) represents the photon, $a'_{\mu}$ (along with its field strength $f'_{\mu \nu}$) represents the hidden photon, and $j_{EM}^\mu$ is the electromagnetic current. In this basis, $a_{\mu}$ couples to the electromagnetic current $j_{EM}^{\mu}$ with coupling strength $e$, while $a'_{\mu}$ is massive and couples to the Standard Model fields only through the kinetic mixing term. The kinetic mixing $\varepsilon$ between $a_{\mu}$ and $a'_{\mu}$ leads to effective interactions between $a'_{\mu}$ and electromagnetic currents, permitting the possibility of discovering the hidden photon.

The mechanisms available for setting the mass and coupling of a hidden photon are highly unrestrictive.
For example, the mass may be set by dimensional transmutation in an asymptotically free sector, allowing it essentially to be arbitrarily small, while the kinetic mixing may be suppressed by loops, powers of the GUT-breaking scale, or small coupling constants.
It is therefore theoretically well motivated to consider hidden photons over an enormous range of mass scales and couplings.
While light hidden photons can be associated with other light hidden-sector states, with interesting experimental consequences of their own, the existence and coupling strength of such states are independent of that of the hidden photon itself.
There is therefore a strong case for developing direct probes of hidden photons.

A significant effort is ongoing to search for hidden photons in a mass range $\mathcal O$(MeV-GeV), using low energy collider and fixed target experiments (see e.g.~\cite{Reece:2009un, Batell:2009di, Bjorken:2009mm, Aubert:2009af, deNiverville:2011it, Hewett:2012ns, Dharmapalan:2012xp}).
Efforts to search for much lighter hidden photons with smaller-scale lab experiments~\cite{Ahlers:2007rd, Jaeckel:2007ch, Caspers:2009cj, Povey:2010hs, Wagner:2010mi, Betz:2013dza, Ehret:2010mh, Arias:2010bh, Horns:2012jf, Ringwald:2012hr, Seviour:2014dqa} have received somewhat less attention. These experiments have the potential to discover hidden photons over a vast range of parameter space, extending from $\mathcal O$(eV) down to, as we shall show, scales as low as $\mathcal O(10^{-18}$eV).

It is well known \cite{Holdom:1985ag} that in the limit where the mass $\mhidden$ of the hidden photon goes to zero, the couplings of the hidden photon to electromagnetic currents can be rotated away. Hence, the physical effects of this hidden photon on the electromagnetic currents, with which we may detect them, have to be proportional to some power of the mass $\mhidden$. This is also true for constraints on these particles that arise as a result of their interactions with electromagnetic currents. The strongest constraints on these scenarios are imposed by astrophysical observations, where the production of these particles from electromagnetic currents in red giants and supernovae can lead to enhanced cooling of such objects. The strength of such bounds is always proportional to a power of $\mhidden$.

Recent work~\cite{An:2013yfc, An:2013yua, Redondo:2013lna} has shown that the
previously calculated bounds on such particles were incomplete.
In earlier work~\cite{Redondo:2008aa}, the cooling rate of stars by hidden-photon emission had been calculated and found to be proportional to
$\varepsilon^2 (\mhidden/T)^4$.
However, as was shown in~\cite{An:2013yfc}, this calculation had neglected the longitudinal polarization of the hidden photon, which exists because the hidden photon has a non-zero mass. Inclusion of the missing longitudinal polarization was found to give a parametrically larger cooling rate, proportional to
$\varepsilon^2 (\mhidden/T)^2$.
This observation significantly strengthened bounds on such particles, dramatically shrinking the parameter space
accessible to a variety of ``Light-Shining-Through-A-Wall" experiments such as ALPS~\cite{Ehret:2009sq, Bahre:2013ywa}.
This reduction in the reach occurs because the design of Light-Shining-Through-A-Wall experiments has been based on the principle of production and subsequent detection of the transverse modes of the
hidden photon.
The transverse modes are, essentially, parametrically more weakly coupled to Standard Model particles than the longitudinal mode that is responsible for the astrophysical bounds.

However, if a longitudinal mode can be emitted in a stellar environment, leading to strong constraints, it should also be possible to utilize it in the laboratory for detection.
In this paper, we will show that the longitudinal mode has qualitatively new behavior in Light-Shining-Through-A-Wall experiments, that can lead to a parametrically stronger signal than that from the transverse modes (proportional to $\varepsilon^2 \mhidden^2$ rather than $\varepsilon^2 \mhidden^4$). This allows a great enhancement of the experimental sensitivity (Fig.~\ref{fig:reach}).
To take full advantage of this sensitivity enhancement, the currently used experimental setups may need simple modifications (Fig.~\ref{fig:setup}).
We begin in section \ref{sec-summary} with a conceptual overview of these qualitatively new effects.
Following this overview, in section \ref{sec-math} we present the governing equations of electromagnetism with a hidden-photon.
Using these, in section \ref{sec:warm-up} we study a mock-up of a Light-Shining-Through-A-Wall experiment using infinite plane-waves, showing how the signal is parametrically enhanced by using longitudinal rather than transverse waves.
In \ref{sec:cavities}, we then turn to the more complicated calculation of the signal in a realistic Light-Shining-Through-A-Wall experiment, of the type proposed in~\cite{Jaeckel:2007ch} using resonant microwave cavities. We show that the parametric enhancement persists, and give explicit formulae for calculating the signal in an arbitrary experimental setup.
We present the implications of our work, both on existing experimental results and for future experiments, in section \ref{sec:consequences}. Finally, we conclude in section \ref{sec:conclusions}.

\section{Conceptual Overview}
\label{sec-summary}

\begin{SCfigure}
\centering
\includegraphics[width=0.55\textwidth]{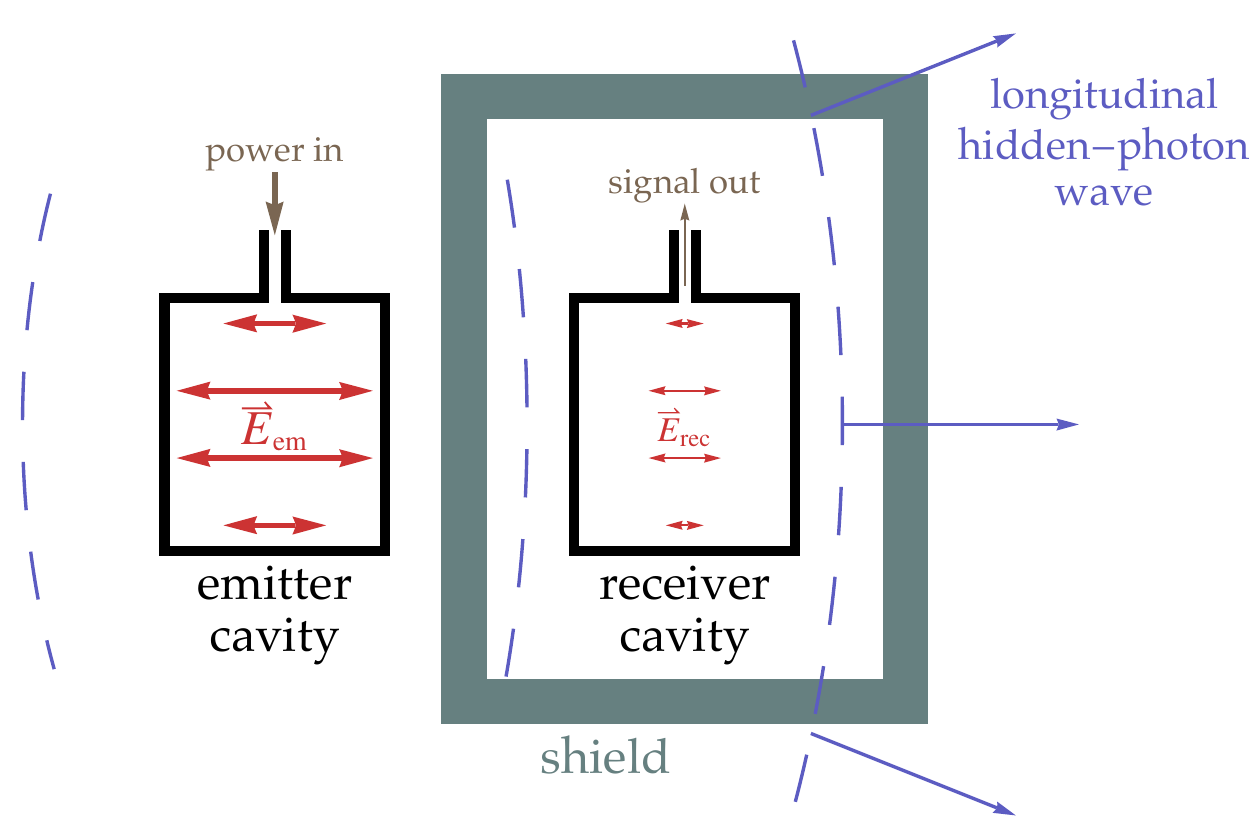}
\caption{Schematic setup of a microwave cavity search for hidden photons, designed to take advantage of the improved transmission of longitudinal hidden-photon waves, and give the largest possible signal field $\vec E_{rec} \sim \varepsilon^2 m_{\gamma'}^2/\omega^2 \vec E_{em}$.
Optimizing the sensitivity requires only a slight modification of the setup proposed in~\cite{Jaeckel:2007ch, Caspers:2009cj}:
\emph{the $\vec E$-field of the driven emitter-cavity mode should point coherently in the direction of the receiver cavity}.
\vspace{10pt}}
\label{fig:setup}
\end{SCfigure}

Hidden photons can be probed by producing and detecting them in the laboratory through their coupling to charge currents.
A classic way to perform such an experiment is to create a source of electromagnetic fields.
The charge current that produces this electromagnetic field will also source hidden photon fields.
A detector located inside an electromagnetic shield is then placed in the vicinity of this source.
The shield will block regular electromagnetic fields from the photon, but permit the weakly coupled hidden photon to leak through, exciting the detector.
Since electromagnetic fields can be efficiently produced, shielded and detected over a wide range of frequencies ranging from the radio to the optical, these ``Light-Shining-Through-A-Wall" experiments are an effective way to probe hidden photons.

Let us first understand the parametric behavior of the signal in these experiments.
Upon diagonalizing the kinetic terms of \eqref{KineticMixing}, we obtain
\begin{equation}
\mathcal L = -\frac{1}{4} \left( F_{\mu \nu} F^{\mu \nu} + F'_{\mu
\nu} F'^{\mu \nu} \right) + \frac{1}{2}\mhidden^2 A'_{\mu} A'^{\mu}
- e J_{EM}^{\mu} \left( \, A_{\mu} + \varepsilon \, A'_{\mu}\right)
\label{eq:mass-basis-Lagrangian}
\end{equation}
where $A_{\mu}$ and $A'_{\mu}$ are of course linear combinations of $a_{\mu}$ and $a'_{\mu}$.
The propagating mass eigenstates are the transverse photon and hidden-photon modes $ |A_T\rangle$ and $ |A'_T \rangle$, along with the longitudinal hidden-photon mode $ |A'_L \rangle$ (there is no longitudinal mode for the massless photon).
In this basis, the linear combination $ |A_T\rangle + \varepsilon |A'_T\rangle$ directly interacts with charges, while the linear combination $ |A'_T\rangle - \varepsilon |A_T\rangle$ is sterile.

Let us first focus on the transverse modes.  In a ``Light-Shining-Through-A-Wall" experiment, the linear combination $|A_T\rangle + \varepsilon |A'_T\rangle$ is first produced.
But, this is not a mass eigenstate - the two states $|A_T\rangle$ and $|A'_T\rangle$ have different masses leading to a differential phase developing between them.
These states are relativistic and hence after traveling a distance $L$ (where the wall is placed), the state evolves to
$|A_T\rangle + \varepsilon \, e^{-i \frac{1}{2} (\mhidden^2/ \omega) L} |A'_T\rangle$ (up to an overall irrelevant phase), where $\omega$ is the energy of the produced state.
Notice that this is similar to the phenomenon of neutrino oscillations.
The linear combination $|A_T\rangle + \varepsilon |A'_T\rangle$ is absorbed by the wall, while the sterile component
$|A'_T\rangle - \varepsilon |A_T\rangle$ passes through the wall.
The amplitude of this sterile component is proportional to $\varepsilon \mhidden^2 L / \omega$.
While this sterile component travels through the wall, it cannot directly be detected by the instruments on the other side of the wall. For detection, we need the sterile state
$ |A'_T\rangle - \varepsilon |A_T\rangle$ to partially oscillate back into the state
$ |A_T\rangle + \varepsilon |A'_T\rangle$.
This will happen because the sterile state is also not a mass eigenstate. The component of the state that overlaps with the interaction state $|A_T\rangle + \varepsilon |A'_T\rangle$ after traveling a further distance $L$ is again proportional to $\varepsilon \mhidden^2 L / \omega$. Hence, the signal in this setup is proportional to $\varepsilon^2 \mhidden^4 L^2 / \omega^2$. For an apparatus of fixed physical size $L$, this implies that the sensitivity for low mass hidden photons ($\mhidden \ll 1/L$) drops sharply. With this scaling the unconstrained parameter space that can be probed by these experiments is limited.

\begin{figure}[t]
\centering
\includegraphics[width=0.7\textwidth]{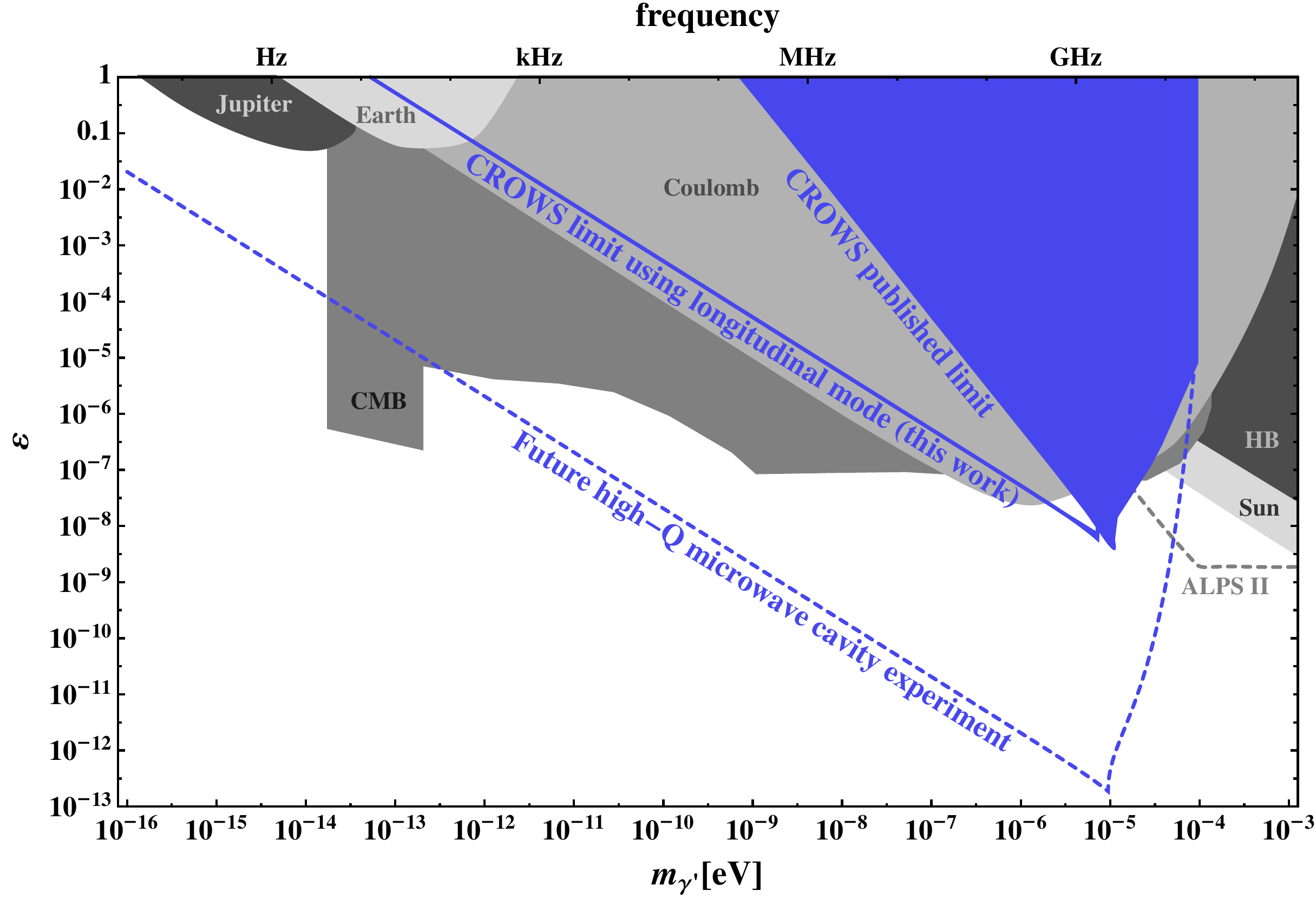}
\caption{The reach of cavity searches for hidden photons, taking advantage of the improved transmission of the longitudinal modes. The solid blue region is the published limit from the CROWS experiment~\cite{Betz:2013dza}, calculated using the results of~\cite{Jaeckel:2007ch}. The solid blue line shows the bound we obtain by reanalyzing the CROWS results, taking into account longitudinal hidden-photon modes. The dashed blue line shows the reach of a realistic future cavity experiment. See section~\ref{sec:consequences} for more details. The gray shaded regions show various preexisting astrophysical and laboratory constraints, compiled from Refs.~\cite{Jaeckel:2010ni, An:2013yfc}, while the gray dashed line shows the projected reach of the ALPS-II experiment~\cite{Bahre:2013ywa}.}
\label{fig:reach}
\end{figure}

Naively, one may think that this discussion should also apply to the longitudinal mode of the hidden photon. There is however a very important difference - the massless photon $A_{\mu}$ does not have a longitudinal mode. Since the hidden photon is coupled to electromagnetic currents (see Eq.~\ref{eq:mass-basis-Lagrangian}), these currents will source its longitudinal mode at order $\varepsilon$. This longitudinal mode will not be blocked by the wall since it is  very weakly coupled.  As the longitudinal mode enters the shield, it will move charges in the shield. This motion will excite photon modes that will oppose the incident longitudinal mode. Since the detector couples to a linear combination of both these fields, one may worry that the photon modes will exactly cancel the linear combination that couples to the detector.
However, such a cancellation of the effects of the longitudinal mode would require another longitudinal mode.
Since the photon does not have such modes in the vacuum, this cancellation cannot occur.

The mode will thus go through the wall, and since it is directly coupled to electromagnetic currents, it can be  detected on the other side.
In this setup, the dependence on $\mhidden$ appears because both the production amplitude of longitudinal modes, and the strength of their effect on electric charges, scales as\footnote{The Goldstone boson equivalence theorem implies that, at lowest order in $m_{\gamma'}^2$, the coupling to longitudinal modes through the $A'_{\mu} J_{EM}^{\mu}$ term is equivalent to coupling to Goldstone bosons through the derivative coupling $1/m_{\gamma'} \partial_{\mu} \phi J_{EM}^{\mu}$.
However, this coupling vanishes identically due to electromagnetic current conservation.
The leading order effect therefore appears at the next order in $m_{\gamma'}^2$, and is suppressed by $m_{\gamma'}/\omega$.} $\mhidden/\omega$.
Hence, if the longitudinal mode is utilized, the signal in the experiment would scale as $\varepsilon^2 \mhidden^2/\omega^2$, more favorably than the transverse mode. This makes such experiments capable of covering significant new parameter space beyond current bounds (see Fig.~\ref{fig:reach}).

This is the main message of this paper.
In the following sections, we will show that the longitudinal modes of the hidden photon can be produced and detected in the laboratory through the scheme described above.
For concreteness, we will focus on Light-Shining-Through-A-Wall experiments using microwave cavities, as proposed in~\cite{Jaeckel:2007ch}. The hidden photon fields produced by driving a cavity can be detected through the excitations of another cavity placed inside a shield (see Fig.~\ref{fig:setup}). This scheme benefits from the fact that large electromagnetic fields can be produced and sustained for significant durations through the use of resonant superconducting cavities. When the receiver cavity is resonantly matched to the source cavity, the signal in the setup can also be resonantly enhanced. While early-stage experiments of this type have already been performed~\cite{Povey:2010hs, Wagner:2010mi, Betz:2013dza}, they were focussed on the detection of the transverse modes of the hidden photon~\cite{Ahlers:2007rd, Jaeckel:2007ch}.

We show that the sensitivity of this setup can be parametrically enhanced if the setup is optimized to detect the longitudinal mode.
Such optimization is necessary since the longitudinal modes are most strongly produced in a direction perpendicular to that of the transverse modes.
The detector cavity would therefore have to be appropriately located  (see Fig.~\ref{fig:setup}) in order for it to most efficiently couple to the longitudinal mode.
Using these results, we will re-examine bounds on hidden photons that have already been placed by existing microwave cavity experiments such as CROWS.
We will show that with their existing results they have already placed parametrically stronger bounds on hidden photons than their published limits (see Fig.~\ref{fig:reach}).
We also briefly discuss (in section~\ref{sec:future-sensitivity}) the potential reach of a proposed superconducting, high-$Q$ microwave cavity search tailored to detect these longitudinal modes~\cite{Sami}. The projected sensitivity of such an experiment is also plotted in Fig.~\ref{fig:reach}.

Unlike microwave cavity experiments, where the utilization of the longitudinal mode parametrically enhances the reach of the experiment into hitherto unexplored regions of parameter space, we do not find a comparable improvement in experiments such as ALPS that utilize optical cavities.
The principal reason for this is that, in the language of the discussion above, the optical experiments are
designed to gain from a large oscillation length $L$.
Although it might be possible to reconfigure such an experiment to use the longitudinal mode, the large length would no longer enhance the signal, and the resulting reach would be uninteresting.
These considerations are of course different for microwave cavities, where the size and separation of the cavities are comparable to the wavelength of the light. We discuss this further in section~\ref{sec:ALPS}.

While optical experiments likely will not benefit from the longitudinal mode, we find that microwave cavities can benefit enormously.  Such an experiment designed with an optimal geometry to take advantage of the longitudinal mode could push the reach into hidden-photon parameter space by many orders of magnitude and exploring interesting parameter space over a mass range $\sim 10^{-18}$ eV to $10^{-4}$ eV.  Thus, we find that microwave cavity experiments can be an extremely powerful probe of new hidden sectors.

\section{Governing Equations}
\label{sec-math}


The equations of motion for the photon and hidden-photon fields, in the mass basis, follow from Eq.~\ref{eq:mass-basis-Lagrangian}, and are given by
\begin{align}
(\partial_t^2 - \nabla^2) V &= \varrho_{EM} &
(\partial_t^2 - \nabla^2) \vec A &= \vec \jmath_{EM}  &
\dot{\vec V} + \nabla \cdot \vec A &= 0 \, ,
\label{eq:A-eom}\\
(\partial_t^2 - \nabla^2 + m_{\gamma'}^2) V' &= \varepsilon \varrho_{EM} &
(\partial_t^2 - \nabla^2 + m_{\gamma'}^2) \vec A' &= \varepsilon \vec \jmath_{EM}  &
\dot{\vec V}' + \nabla \cdot \vec A' &= 0 \, .
\label{eq:A'-eom}
\end{align}
We will work in the mass basis (rather than the interaction basis) for the duration of this paper\footnote{The final results are, of course, independent of the basis choice. We prefer the mass basis because it avoids the physical longitudinal hidden-photon mode being spread between the $A_\mu$ and  $A_\mu'$ fields. It also preserves the usual form of electromagnetic gauge invariance, with the gauge transformation acting only on the $A_\mu$ field.}.
Eqs.~\ref{eq:A-eom} are simply Maxwell's equations\footnote{
In our conventions, $\vec E^(\vphantom{E}'^)$ and $\vec B^(\vphantom{E}'^)$ are given by $\vec E^(\vphantom{E}'^) = - \vec \nabla V^(\vphantom{E}'^) - \dot{\vec A}^(\vphantom{E}'^) \, ; \quad \vec B^(\vphantom{E}'^) = \vec \nabla \times \vec A^(\vphantom{E}'^)$}
 (with the convenient choice of Lorenz gauge).
Eqs.~\ref{eq:A'-eom} are the Proca equations for a massive vector. They show that the massive hidden-photon field is sourced by electric charge density $\varrho_{EM}$ and current $\vec \jmath_{EM}$ in the same way as the massless photon field, but suppressed by a factor of $\varepsilon$.
The final part of Eqs.~\ref{eq:A'-eom} is a constraint equivalent to conservation of electric charge.

Any setup that produces ordinary electric and magnetic fields will, through Eqs.~\ref{eq:A'-eom}, also source hidden-photon fields at $\mathcal O(\varepsilon)$.
In turn, the Lorentz force on charged particles receives an $\varepsilon$-suppressed contribution from these hidden-photon fields,
\begin{equation}
\vec F = q \Big[\big(\vec E + \varepsilon \vec E' \big) + \vec v \times \big(\vec B + \varepsilon \vec B' \big)\Big] \qquad \text{(modified Lorentz force)}\, .
\label{eq:Lorentz-force}
\end{equation}
This generates small modifications to the electromagnetic currents, and consequently to the electric and magnetic fields, at $\mathcal O(\varepsilon^2)$.
In particular, at the surface of a perfect conductor, Eq.~\ref{eq:Lorentz-force} implies that the electric field obeys the modified boundary condition
\begin{equation}
(\vec E + \varepsilon \vec E')_\parallel = 0 \qquad \text{(B.C. for a conducting surface)} \, ,
\label{eq:conductor-BC}
\end{equation}
which ensures that conduction electrons experience no force parallel to the conductor's surface. (See appendix \ref{sec:skin-depth} for further discussion of why this boundary condition is correct.)

\subsection{Example: electric dipole radiation}
\label{sec:oscillating-edm}

We now consider an instructive example: the transverse and longitudinal waves radiated from an oscillating electric dipole.
In complex notation, an electric dipole aligned along the $\hat z$ axis  can be described by $\vec \jmath_{EM}(\vec x) = j_0 \delta^3(\vec x) e^{i \omega t} \, \hat z$ (along with a corresponding charge density).
$\vec A$ and $\vec A'$ are easily found by solving Eqs.~\ref{eq:A-eom} and~\ref{eq:A'-eom}, giving
\begin{gather}
\vec A(\vec r, t) = \frac{j_0}{4\pi r} e^{i(\omega t - \omega r)} \, \hat z
\qquad\qquad
\vec A'(\vec r, t) = \frac{\varepsilon j_0}{4\pi r} e^{i(\omega t - k r)} \,  \hat z \, ,
\end{gather}
where $k^2 = \omega^2 - m_{\gamma'}^2$. $V$ and $V'$ are given by
\begin{equation}
V(z, t) = \frac{i}{\omega} \vec \nabla \cdot \vec A\,(z, t)
\qquad \qquad
V'(z, t) = \frac{i}{\omega} \vec \nabla \cdot \vec A'(z, t) \, .
\end{equation}
Radiation of transverse and longitudinal modes can been seen from the $\mathcal O(1/r)$ terms in $\vec E$ and $\vec E'$, 
\begin{align}
\vec E(\vec r, t) &=
- \vec \nabla V - i \omega \vec A =
 i \frac{j_0 \omega}{4 \pi r} e^{i (\omega t - \omega r)} \sin \theta \, \hat \theta + \mathcal O \Big(\frac{1}{r^2}\Big)\\
\vec E'(\vec r, t) &=
- \vec \nabla V' - i \omega \vec A' =
 i \varepsilon \frac{j_0 \omega}{4 \pi r} e^{i (\omega t - k r)} \Big(\sin \theta \, \hat \theta - \frac{m_{\gamma'}^2}{\omega^2} \cos \theta \, \hat r \Big) + \mathcal O \Big(\frac{1}{r^2}\Big) \, .
\label{eq:edm-E'}
\end{align}

Radiation of the massless photon is, of course, purely transverse ($\vec E \propto \hat \theta$), and strongest in the  directions perpendicular to the dipole's axis. Radiation of the massive hidden photon also has a longitudinal component ($\propto \hat r$), which is suppressed by a factor $(m_{\gamma'}/\omega)^2$ relative to the transverse mode. Fig.~\ref{fig:edm-radiation-pattern} shows the radiation pattern of $|E'_{\theta, r}|^2$. In contrast to standard dipole radiation, the longitudinal mode is radiated most strongly \emph{along the dipole's axis}.

\section{Toy calculation: screened plane waves}
\label{sec:warm-up}

To demonstrate the difference in screening between transverse and longitudinal modes, we begin with a very simple mockup of a general Light-Shining-Through-A-Wall experiment.
Plane waves of the photon and hidden photon fields are radiated from an infinite sheet of oscillating currents. The waves are screened by a nearby sheet of perfectly conducting metal, and the net force on a test charge is measured on the far side.
Taking in turn transverse and then longitudinal waves, we will see that the force is suppressed by a factor of
$\varepsilon^2 (m_{\gamma'}/\omega)^4$ in the transverse case, but
only $\varepsilon^2 (m_{\gamma'}/\omega)^2$ in the longitudinal one.


\subsection{Generating transverse and longitudinal plane waves}
\label{sec:GeneratePW}

Imagine that at $z=0$ we have an infinite plane of oscillating electric charges. The plane is charge-neutral, but there is a non-zero net current
\begin{eqnarray}
\vec{\jmath}\,(\vec x, t) = \vec \jmath_0 \, \delta(z) e^{i\omega t} \, .
\label{Eq:TransCurrent}
\end{eqnarray}
Depending on whether $\vec \jmath_0$ points along or out of the $x$-$y$ plane, this oscillating current will radiate either transverse or longitudinal waves of the photon and hidden photon fields.

The radiated fields are found by solving Eqs.~\ref{eq:A-eom} and~\ref{eq:A'-eom}:
\begin{align}
(\partial_z^2 + \omega^2) \vec A\,(z, t) = -\vec \jmath_0 \, \delta(z) e^{i\omega t}
&&
(\partial_z^2 + \omega^2 - m_{\gamma'}^2) \vec A'(z, t) = -\varepsilon \vec \jmath_0 \, \delta(z) e^{i\omega t}
& \\
V(z, t) = \frac{i}{\omega} \vec \nabla \cdot \vec A\,(z, t)
&&
V'(z, t) = \frac{i}{\omega} \vec \nabla \cdot \vec A'(z, t) \quad
& \, ,
\label{eq:constraints}
\end{align}
where $k^2 = \omega^2 - m_{\gamma'}^2$.
These equations are easily solved to give the radiated $\vec E$, $\vec B$, $\vec E'$ and $\vec B'$ fields. If the oscillating current points along the plane, we have
\begin{align}
\vec \jmath_0 = j_0 \, \hat x \, : &&
\left\{ \begin{aligned}
  \mymatrix{ \vec E \, \\ \vec B } &= -\frac{j_0}{2} \, e^{i(\omega t - \omega |z|)} \, \mymatrix{ \hat{x} \\ \pm \hat y } \\
  \mymatrix{ \vec E' \\ \vec B' } &= -\varepsilon \frac{j_0}{2} \, e^{i(\omega t - k |z|)} \, \mymatrix{ \frac{\omega}{k} \, \hat{x} \\ \pm \, \hat y }
\end{aligned} \right\}
&& \text{(transverse radiation)} \, .
\label{eq:transverse-plane-radiation}
\end{align}
If the oscillating current points out of the plane, we have
\begin{align}
\vec \jmath_0 = j_0 \, \hat z \, : &&
\left\{ \begin{aligned}
  \mymatrix{ \vec E \, \\ \vec B } &= 0 \\
 \mymatrix{ \vec E' \\ \vec B' } &= -\varepsilon \frac{m_{\gamma'}^2}{\omega k} \frac{j_0}{2} \, e^{i(\omega t - k |z|)} \, \mymatrix{ \, \hat{z} \, \\ 0 }
\end{aligned} \right\}
&& \text{(longitudinal radiation)} \, . \hspace{-10pt}
\label{eq:plane-longitudinal-radiation}
\end{align}

As required by gauge invariance, there is no longitudinal wave of the massless photon field. Here this is enforced by the gauge condition $\partial_\mu A^\mu=0$, combined with the fact that the photon's wavevector is the same as its frequency. However, $k \neq \omega$ for the hidden-photon field, allowing a longitudinal $\vec E'$ wave to be generated at order $\varepsilon m_{\gamma'}^2 / \omega^2$.

\subsection{Screening transverse plane waves}

We will now see how a perfectly conducting thin wall, placed at $z=L$,
affects to the incoming plane waves.  Let us first focus on the
transverse modes, which are generated when $\vec{\jmath} \propto \hat x$.

The incoming plane waves act on the conduction electrons in the wall.
In response these electrons oscillate, causing further transverse waves to be radiated from the wall.
These waves have the same form as Eq.~\ref{eq:transverse-plane-radiation}, but with $z$ replaced by $z-L$.
Because the wall is a perfect conductor, these radiated fields are exactly those needed to cancel the net force on the electrons parallel to the wall. In other words, we have a boundary condition at the wall given by
\begin{equation}
(E_x + \varepsilon E'_x ) \big|_{z = L} = 0 \, ,
\end{equation}
which is a slight modification of the usual boundary condition at a conducting surface.\footnote{For completeness, we confirm this in appendix~\ref{sec:skin-depth} with a more careful look at the propagation of waves in a conductor.} This fixes the full solution to be
\begin{align}
\vec E (z, t) &= - \frac{j_0}{2} \, e^{i\omega t } \, \hat{x} \big(e^{-i \omega |z|} + c \, e^{-i \omega |z-L|} \big)
\\
\vec E' (z, t) &= - \varepsilon \frac{\omega}{k} \frac{j_0}{2} \, e^{i\omega t} \, \hat{x} \big( e^{-i k |z|}
+ c \, e^{-i k |z-L|} \big)
\\
c &= - e^{-i \omega L}\Big(1+\varepsilon^2 \frac{\omega}{k} (e^{i(\omega-k)L} - 1) \Big) +\mathcal O(\varepsilon^4) \, .
\end{align}
On the far side of the wall ($z>L$), the force on a test charge $q$ is then given by
\begin{equation}
\vec F_{\rm probe} = q(\vec E + \varepsilon \vec E') = \frac{\omega}{k} \frac{j_0}{2} e^{i(\omega t - \omega z)}
\, \overbrace{\varepsilon(e^{i(\omega-k)L} - 1)}^{\mathclap{\text{generate }\vec E' - \varepsilon \vec E \text{ at wall}}}
\, \underbrace{\varepsilon(1-e^{i(\omega-k)(z-L)})}_{\mathclap{\text{regenerate }\vec E + \varepsilon \vec E' \text{ beyond wall}}} \, \hat x \, .
\end{equation}
Let us understand this equation in another way. The currents at $z=0$ source the field combination $\vec E + \varepsilon \vec E'$, and this same combination is blocked by the wall. In order for any field to penetrate the wall, $\vec E'$ must become out of phase with $\vec E$, giving a non-zero $\vec E' \!- \varepsilon \vec E$ at the wall (at order $\varepsilon$). The first term in parentheses is just this phase difference. In order to exert a force on a test charge beyond the wall, a non-zero $\vec E + \varepsilon \vec E'$ must be regenerated (at one higher order in $\varepsilon$) by a further relative phase shift. This second phase shift corresponds to the second term in parentheses. Both of these relative phase shifts occur because of the finite hidden-photon mass, and vanish as $m_{\gamma'}^2$ in the small $m_{\gamma'}$ limit.

In the small mass limit ($m_{\gamma'}^2 \ll \omega / L $), we then have
\begin{equation}
\vec F_{\rm probe} \longrightarrow q\varepsilon^2 \frac{m_{\gamma'}^4}{\omega^4} \frac{\omega^2 L(z-L) j_0}{8} e^{i(\omega t - \omega z)} \, \hat x \, ,
\end{equation}
and we see that the conducting wall is effective at screening transverse radiation up to order $\varepsilon^2 m_{\gamma'}^4 / \omega^4$.

\subsection{Screening longitudinal plane waves}

We now turn to effect of the conducting wall on the longitudinal hidden-photon mode, radiated when $\vec \jmath \propto \hat z$.
The conduction electrons in the wall respond to the small force they feel from the incoming wave. This causes them to oscillate in the $\hat z$ direction at order $\varepsilon^2$. However, since there is no longitudinal mode for the massless photon, they cannot generate any further $\vec E$ field beyond the wall. They do of course radiate a further $\vec E'$ field, but this occurs at order $\varepsilon^3$, and we are not interested in it.
The force on a test charge $q$ beyond the wall then follows directly from Eq.~\ref{eq:plane-longitudinal-radiation},
\begin{align}
\vec F_{\rm probe} &= - q \varepsilon^2 \frac{m_{\gamma'}^2}{\omega k} \frac{j_0}{2} \, e^{i(\omega t - k z)} \, \hat{z} \\
&\xrightarrow{m_{\gamma'}\ll \omega} - q \varepsilon^2 \frac{m_{\gamma'}^2}{\omega^2} \frac{j_0}{2} \, e^{i(\omega t - \omega z)} \, \hat{z} \, .
\end{align}

Altogether, we see that the conducting wall does not have any effect: it simply cannot screen the longitudinal hidden-photon wave.
As a result, the force experienced by the test charge is a factor $\omega^2 / m_{\gamma'}^2$ larger for longitudinal radiation than for transverse radiation in the limit of a light hidden photon.

One might be concerned that these scalings arise partly because of the infinite extent of the plane wave, allowing $1/m_{\gamma'}^2$ factors to appear from a volume integral.
We now turn to realistic Light-Shining-Through-A-Wall experiments using resonant cavities, and will see that these scalings do persist:
the signal from longitudinal waves is indeed parametrically larger than that from transverse waves in the $m_{\gamma'}\ll \omega$ regime.

\section{Signal in a hidden-photon search with resonant cavities}
\label{sec:cavities}

Having gained an understanding of hidden photon behavior in a simplistic setup in section~\ref{sec:warm-up}, we turn to the real setup of interest: a Light-Shining-Through-A-Wall experiment using resonant cavities, of the kind proposed in~\cite{Jaeckel:2007ch} (see Fig.~\ref{fig:setup}).
We begin the section by presenting a general prescription for calculating the signal, which can be easily employed in any given experimental setup.
We follow this in section~\ref{sec:cavity-example} with an explicit, simple example, demonstrating setups that do and do not utilize the improved transmission of the longitudinal hidden-photon mode. A careful derivation of the signal can be found in appendix~\ref{sec:detailed-cavity-derivation}, and a discussion of previous attempts in the literature~\cite{Ahlers:2007rd, Jaeckel:2007ch} in appendix~\ref{sec:previous-literature}.

\subsection{General prescription for an arbitrary experimental geometry}
\label{sec:general-cavity-solution}

Take
$\vec E(\vec r, t)= \vec E_{em}(\vec r) e^{i \omega t}$ and $\vec B(\vec r, t) = \vec B_{em}(\vec r) e^{i \omega t}$
to be the (known) $E$- and $B$-fields of the cavity mode that is driven inside the emitter cavity. The emitter cavity then radiates a hidden-photon field, with the hidden electric field $\vec E'$ given by
\begin{equation}
\vec E'(\vec r, t) = - \varepsilon m_{\gamma'}^2 \Bigg[ \int_{em} \!\!d^3 x \, \frac{\vec E_{em}(\vec x)}{4\pi \left| \vec r - \vec x \right|} \, e^{-i k \left| \vec r - \vec x \right|} \Bigg]  e^{i\omega t}\, .
\label{eq:E'-radiated}
\end{equation}
Here the integral is over the interior of the emitter-cavity, and $k^2 \equiv \omega^2 - m_{\gamma'}^2$. If $\,m_{\gamma'}\!>\!\omega$, the hidden-photon cannot be radiated, but is nonetheless sourced within a distance $\sim1/m_{\gamma'}$ of the emitter cavity. In this case Eq.~\ref{eq:E'-radiated} is still valid after replacing $k$ with $-i \kappa$, where $\kappa^2 \equiv m_{\gamma'}^2 - \omega^2$.

We see that the radiated hidden-photon fields are suppressed by a factor $\varepsilon m_{\gamma'}^2/\omega^2$ relative to the emitter-cavity fields. For the longitudinal mode, this is the same suppression that we saw in Eq.~\ref{eq:edm-E'} for the oscillating electric dipole: the field strength of a longitudinal wave always appears at order $m_{\gamma'}^2$. For the transverse modes, the $m_{\gamma'}^2$ arises because there is perfect destructive interference outside the cavity in the massless limit.

The hidden-photon fields penetrate the receiver cavity, where they excite a resonant response of the matching receiver-cavity mode. After allowing $\sim 2 \pi Q$ cycles for the resonance to ring up, the observed signal fields within the receiver cavity are given by
\footnote{Here we have included the $Q$-factor only of the receiver cavity, treating the emitter cavity for simplicity as a perfect resonator with no linewidth. The true signal would follow straightforwardly by integrating Eq.~\ref{eq:signal-with-full-t-and-omega-dependence} over the emitter-cavity lineshape, but this is relevant only for a detailed signal-processing analysis, which is beyond the scope of this work.}
\begin{align}
\vec E_{\rm observed}(\vec r, t) &= - \frac{Q}{\omega} \Bigg[ \frac{ \int_{rec} d^3 x \, \vec E_{cav}^*(\vec x) \cdot \vec \jmath_{\rm eff} (\vec x) }{ \int_{rec} d^3 x \, |E_{cav}(\vec x)|^2 } \Bigg] \vec E_{cav}(\vec r) e^{i \omega t} \\
\vec B_{\rm observed}(\vec r, t) &= - \frac{Q}{\omega} \Bigg[ \frac{ \int_{rec} d^3 x \, \vec E_{cav}^*(\vec x) \cdot \vec \jmath_{\rm eff} (\vec x) }{ \int_{rec} d^3 x \, |E_{cav}(\vec x)|^2 } \Bigg] \vec B_{cav}(\vec r) e^{i \omega t}
\label{eq:B-signal}\\
\vec \jmath_{\rm eff}(\vec x) &\equiv - \frac{i \varepsilon}{\omega} \Big[m_{\gamma'}^2 \vec E'(\vec x,0) - \vec \nabla \big(\vec \nabla \cdot \vec E'(\vec x,0)\big)\Big]
\label{eq:j_eff}  \, .
\end{align}
Here the integrals are over the interior of the receiver cavity, and $\vec E_{cav}(\vec x)$ and $\vec B_{cav}(\vec x) = i \vec \nabla \times \vec E_{cav}(\vec x)/\omega$ are the (known) spatial $E$- and $B$-field profiles of the excited cavity mode.

The function $\vec \jmath_{\rm eff}(\vec x)$ appearing above deserves further attention, since it captures the key result of this paper. If the radiated hidden-photon field is purely transverse, then by definition $\vec \nabla \cdot \vec E'=0$, whereas if it is purely longitudinal then $ \vec \nabla \big(\vec \nabla \cdot \vec E') = - k^2 \vec E'$. In both cases $\vec \jmath_{\rm eff}$ simplifies, and we can write
\begin{equation}
\vec \jmath_{\rm eff}(\vec x) =
- \frac{i \varepsilon}{\omega} \vec E'(\vec x,0) \times \begin{cases}
 m_{\gamma'}^2 & \text{(Pure transverse)} \\[5pt]
 \omega^2 & \text{(Pure longitudinal)} \, .
\end{cases}
\label{eq:jeff_trans_vs_long}
\end{equation}
Comparing the two cases, we immediately see the parametric enhancement of the signal from the longitudinal mode over the transverse mode in the small mass limit.

Eqs.~\ref{eq:E'-radiated}-\ref{eq:j_eff} provide a complete prescription for calculating the expected signal in the receiver cavity in any experimental setup.
Note that this signal is not affected by shielding or any other material placed between the cavities.
The factors in square brackets are in general to be calculated numerically for a given setup, but in the following subsection we go through a simple example in which the signal can be calculated analytically.

\subsection{Example: well-separated cylindrical cavities}
\label{sec:cavity-example}

\begin{figure}[t]
\centering
\subfigure[~Radiation pattern from an oscillating EDM.]{
  \includegraphics[width=0.45\textwidth]{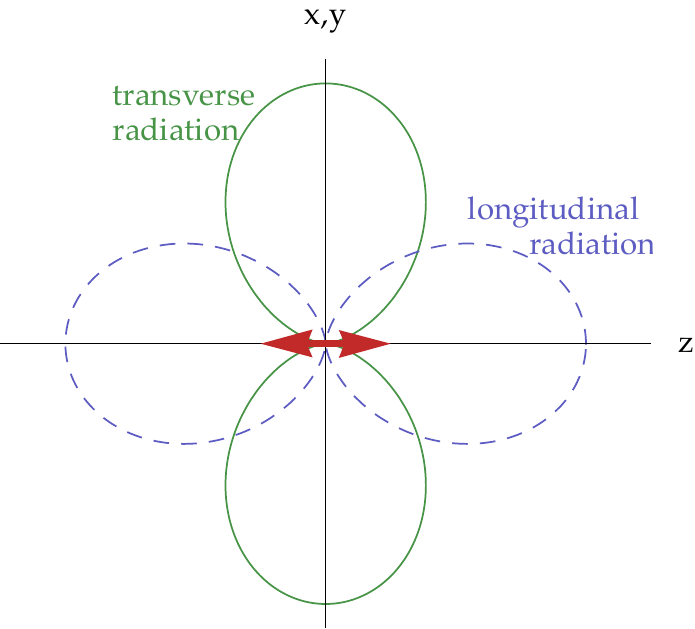}
  \label{fig:edm-radiation-pattern}
  }
\hfill
\subfigure[~Good and bad cavity positioning.]{
  \includegraphics[width=0.48\textwidth]{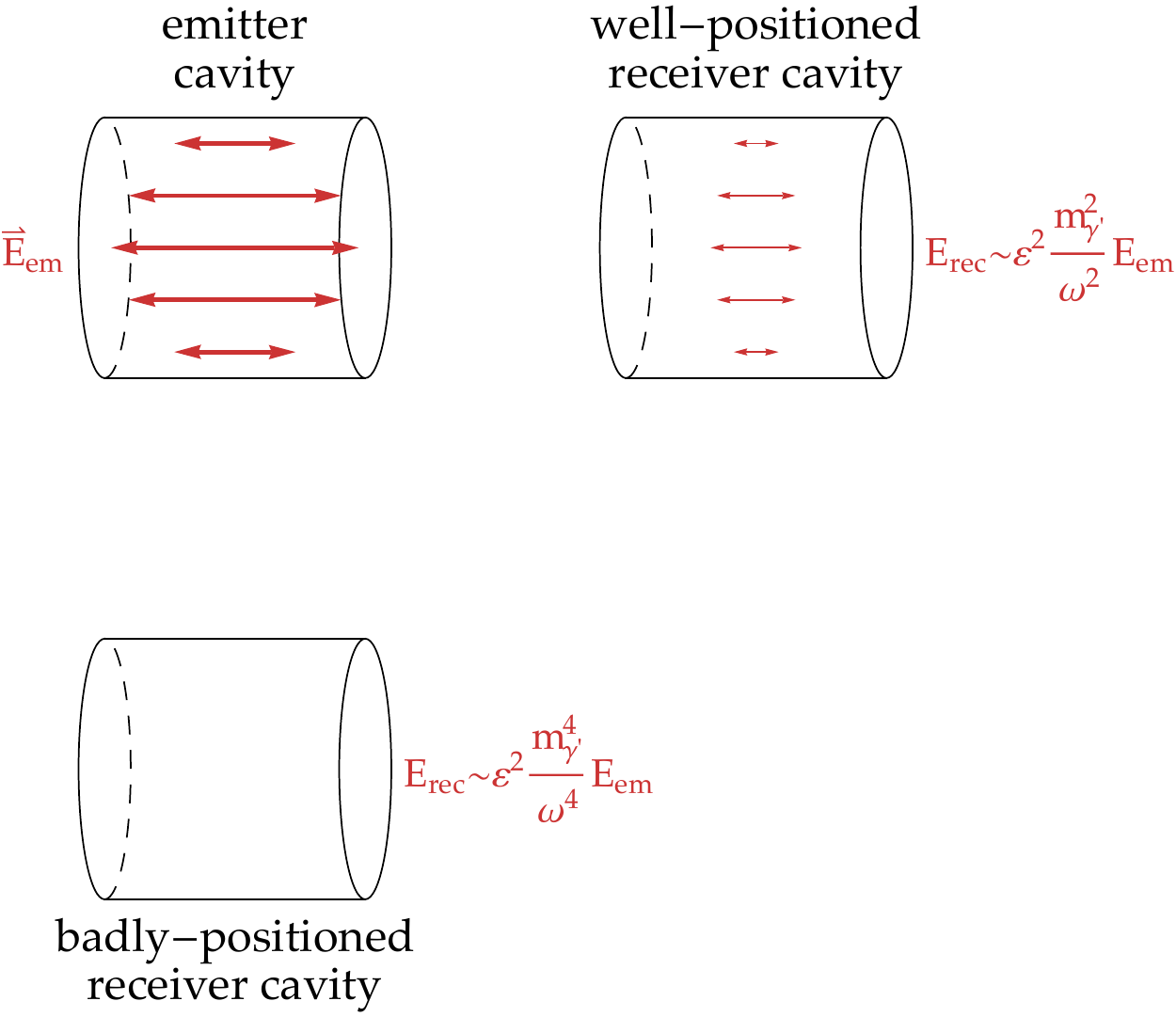}
  \label{fig:cavity-positioning}
  }
\caption{(a) shows the radiation pattern from an oscillating electric dipole, which is similar to that from the emitter cavity shown on the right. Transverse modes are mainly emitted perpendicular to the dipole / emitter~$E$-field, while the longitudinal hidden photon mode is emitted mainly parallel to it.
Figure~\ref{fig:edm-and-setup}: (b) examples of cavity setups that are (upper left + upper right) and are not (upper left + lower left) well designed to efficiently receive the longitudinal mode from the emitter (see sections~\ref{sec:cavity-positioning},~\ref{sec:cavity-example}). The distinctive dependence of the signal on the orientation of the cavities is a useful way to distinguish a true signal from an unexpected field leakage.}
\label{fig:edm-and-setup}
\end{figure}

We now calculate the signal explicitly for a particularly simple setup: well-separated identical cylindrical cavities, in which the cavity mode used is the lowest mode with $E$-field pointing along the cavity's axis (this is known as the TM$_{010}$ mode). We will consider two possible orientations of the cavities, as show in Fig.~\ref{fig:cavity-positioning}: one with the cavities aligned along their cylindrical axes, and another with them separated perpendicular to their axes.
We will see that only the first setup gains the parametric enhancement from the longitudinal hidden-photon mode.
In practice it is better to place the cavities close together, in which case near-field terms become important and the distinction between the two setups is blurred.

Take the cavities to have length $L$ and radius $R$, and to be aligned along the $z$-axis (which points to the right in Fig.~\ref{fig:cavity-positioning}).
At $\mathcal O(\varepsilon^0)$ the fields of the emitter cavity mode are given by
\begin{equation}
\mymatrix{\vec E_{em} \\ \vec B_{em}} = \mymatrix{-i J_0(\omega \rho) \hat z \\  J_1(\omega \rho) \hat \phi } B_{em} \,
\label{eq:emitter-cavity-mode}
\end{equation}
inside the emitter cavity (and vanish outside).
Here we are using cylindrical coordinates $(z, \rho, \phi)$, and $J_{0,1}$ are Bessel functions. The frequency is given by $\omega = \alpha_{0 1}/R$, where $\alpha_{0 1}\approx2.4$ is the first zero of $J_0(z)$.

\subsubsection{Hidden photon emission at $\mathcal O(\varepsilon)$}
Following Eq.~\ref{eq:E'-radiated}, the radiated hidden photon fields are given by
\begin{equation}
\vec E'(\vec r, t) = i \varepsilon m_{\gamma'}^2 B_{em} \, \hat z \int_{L/2}^{L/2} d z \int_0^R \! \rho \, d \rho \int_0^{2 \pi} \!d \phi \frac{J_0(\omega \rho)}{4\pi \left| \vec r - \vec x \right|} e^{i(\omega t - k \left| \vec r - \vec x \right|)} \, .
\label{eq:E'-from-cavity-full}
\end{equation}
In the near-field region this must be solved numerically. However for well-separated cavities ($r \gg R, L$) we can just keep the leading term in $1/r$, with the result
\begin{equation}
\mymatrix{\vec E' \\ \vec B' } = i \varepsilon \frac{m_{\gamma'}^2}{\omega^2} \frac{L}{r} C(\theta) \,  B_{em} \, e^{i(\omega t - k r)} \times \mymatrix{ \hat z \\ -\frac{k}{\omega} \sin \theta \, \hat \phi } + \mathcal O\Big(\frac{1}{r^2}\Big)  \, .
\label{eq:E'-radiated-good-setup}
\end{equation}
The $\mathcal O(1)$ function $C(\theta)$, given by
\begin{equation}
C(\theta) \equiv
\frac{1}{2} \alpha_{01} J_1(\alpha_{01}) \frac{J_0(k R \sin \theta)}{\cos^2 \theta + m_{\gamma'}^2/\omega^2 \sin^2 \theta} \frac{\sin( \frac{1}{2} k L \cos \theta)}{\frac{1}{2} k L \cos \theta} \, ,
\end{equation}
adds a rather mild angular dependence to the fields as long as $L \simlt R$. The radiation pattern of longitudinal and transverse radiation is then similar to that from the oscillating electric dipole considered in section~\ref{sec:oscillating-edm}, and shown in Fig.~\ref{fig:edm-radiation-pattern}, with the longitudinal component $E'_r$ radiated most strongly along the cavity's axis, and transverse component $E'_\theta$ most strongly perpendicular to it.

\subsubsection{Response of a well-positioned receiver cavity: $\mathcal O (\varepsilon^2 m_{\gamma'}^2)$ signal}

Here we will see that if the receiver cavity is well positioned with respect to the emitter, the signal field $\vec B_{observed}$ will be of order $Q \varepsilon^2 m_{\gamma'}^2/\omega^2 B_{em}$. Take the receiver cavity to be placed a distance $d\gg L, R$ \emph{along the direction of the emitter cavity's $\vec E$ field} (i.e. along the $z$-axis). Here the hidden photon field is approximately a longitudinal plane wave,
\begin{equation}
\mymatrix{ \vec E' \\ \vec B' } = \mymatrix{ \,\hat z \, \\ 0 } \, E'_0 \, e^{i(\omega t - k z)}
\, , \qquad
E'_0 =  -  \frac{1}{2} i \varepsilon \frac{m_{\gamma'}^2}{\omega^2} \frac{L}{d} \alpha_{01} J_1(\alpha_{01}) \bigg( \frac{\sin( \frac{1}{2} k L)}{\frac{1}{2} k L} \bigg)   \,  B_{em} \, ,
\label{eq:hidden-photon-fields-above-emitter}
\end{equation}
and
\begin{gather}
\vec \jmath_{\rm eff}(\vec x) =
 - \frac{i \varepsilon}{\omega} (m_{\gamma'}^2 + k^2) \vec E'(\vec x,0) =
 - i \varepsilon \omega  E'_0 \, e^{-i k z} \hat z \, .
 \end{gather}
The hidden-photon excites the same TM$_{010}$ mode as is driven in the emitter cavity, giving
 \begin{align}
\Bigg[ \frac{ \int_{rec} d^3 x \, \vec E_{cav}^*(\vec x) \cdot \vec \jmath_{\rm eff} (\vec x) }{ \int_{rec} d^3 x \, |E_{cav}(\vec x)|^2 } \Bigg]
&= \Bigg[ \frac{ \int_{d-L/2}^{d+L/2} dz \int_0^R \rho \, d \rho \, (i J_0(\omega \rho))  (- i \varepsilon \omega  E'_0 \, e^{-i k z}) }{ L \int_0^R \rho \, d \rho  \, J_0(\omega \rho)^2 } \Bigg] \\
&= \frac{2 \varepsilon \omega }{\alpha_{0 1} J_1(\alpha_{0 1})} \bigg( \frac{\sin (\frac{1}{2} k L)}{\frac{1}{2}k L} \bigg)  E'_0 e^{-i k d}
\end{align}

This gives the final result for the fully-rung-up signal field in the receiver cavity, with the $m_{\gamma'}^2$ scaling as expected:
\begin{gather}
\vec B_{observed} = i J_1(\omega \rho) \hat \phi \, e^{i (\omega t - k d)} \, B_{rec} \\
B_{rec} = Q \, \varepsilon^2 \frac{m_{\gamma'}^2}{\omega^2} \frac{L}{d} \bigg( \frac{\sin (\frac{1}{2} k L)}{\frac{1}{2}k L} \bigg)^{\!2}\, B_{em} \, .
\label{eq:receiver-cavity-signal}
\end{gather}


\subsubsection{Response of a badly positioned receiver cavity: $\mathcal O(\varepsilon^2 m_{\gamma'}^4)$ signal}

For a general arrangement of emitter and receiver cavities, Eq.~\ref{eq:receiver-cavity-signal} will typically give the correct parametric signal size.
However, as we now demonstrate, if the receiver cavity is badly positioned the signal can vanish at order $m_{\gamma'}^2$, resulting in a parametrically smaller signal of $\vec B_{observed} \sim Q \varepsilon^2 m_{\gamma'}^4/\omega^4 B_{em}$.
As we observed earlier, this occurs when the receiver cavity is in a purely transverse $\vec E'$ field.

Take the receiver cavity to now be placed a distance $d\gg L,R$ away from the emitter in the direction \emph{perpendicular} to the $\vec E$ field (along the $x$-axis). The hidden photon field here is approximately a transverse plane wave,
\begin{equation}
\mymatrix{\vec E' \\ \vec B'} = E'_0 \, e^{i(\omega t - k x)} \, \mymatrix{ \hat z \\ - \frac{k}{\omega} \hat y }
\, , \qquad
E'_0 =  - \frac{1}{2} i \varepsilon \frac{L}{d} \alpha_{01} J_1(\alpha_{01}) J_0(k R)   \,  B_{em} \, .
\end{equation}
Now $\vec \nabla \cdot \vec E'=0\,$, and so $\vec \jmath_{\rm eff}$ is suppressed by $m_{\gamma'}^2$:
\begin{gather}
\vec \jmath_{\rm eff}(\vec x) =
 - \frac{i \varepsilon}{\omega} m_{\gamma'}^2 \vec E'(\vec x,0) =
 - i  \frac{m_{\gamma'}^2}{\omega^2} \varepsilon \omega  E'_0 \, e^{-i k x} \hat z \, .
 \end{gather}

Again evaluating Eq.~\ref{eq:B-signal}, we now find that an $m_{\gamma'}^4$ suppressed field is generated in the receiver cavity,
\begin{gather}
\vec B_{observed} = i J_1(\omega \rho) \hat \phi \, B_{rec} \, e^{i (\omega t-k d)} \\
\begin{aligned}
B_{rec} = Q \, \varepsilon^2 \frac{L}{d} \,\Big( J_0(\alpha_{01}k /\omega) \Big)^2 \, B_{em}
\xrightarrow{m_{\gamma'}\ll \omega} 0.39 \, Q \, \varepsilon^2 \frac{m_{\gamma'}^4}{\omega^4} \frac{L}{d} \, B_{em} \, .
\end{aligned}
\label{eq:bad-receiver-cavity-signal}
\end{gather}


\section{Consequences for existing and future experiments}
\label{sec:consequences}

\subsection{Cavity positioning}
\label{sec:cavity-positioning}

It should be clear from the examples in section~\ref{sec:cavity-example} that the positioning of the cavities and the choice of cavity mode can have a significant effect on the size of the signal, with the wrong setup resulting in a parametrically weaker reach.
The examples above were greatly simplified by taking the large separation limit and simple cylindrical geometry:
a real experiment with cavities placed close together will require a numerical study to determine the optimum setup.
We emphasize that \emph{the optimum setup will be different from those previously chosen}, which were based on the signal from transverse modes only presented in Ref.~\cite{Jaeckel:2007ch}.

In addition, due to the distinctive angular dependence of the receiver-cavity response, measuring the variation of the signal with cavity orientation should provide an important way to distinguish it from an unknown background.

\subsection{Limit from the CROWS experiment}

A search for hidden photons with high-$Q$ resonant cavities has already been carried out by the CERN Resonant Weakly Interacting
sub-eV Particle Search (CROWS) experiment~\cite{Betz:2013dza}.
The optimal setup to employ, and the results obtained, were determined using the calculation of Ref.~\cite{Jaeckel:2007ch}
-- that is to say, accounting only for transverse hidden-photon modes.
Consequently, the chosen cavity mode was not optimized for the emission and detection of longitudinal modes
(specifically, the chosen mode was TE$_{011}$, which emits no longitudinal hidden-photon radiation in the far-field limit).

However, CROWS also carried out a search for Axion-Like Particles, with the addition of a static $B$-field through the cavities and a different choice of cavity mode. This setup was coincidently ideal for producing and detecting longitudinal hidden-photon radiation (it used the TM$_{010}$ mode described above, and the ``good cavity positioning'' illustrated in Fig.~\ref{fig:cavity-positioning}).
In this section we reinterpret the results of this search to place new limits on hidden photons, taking advantage of the improved penetration of longitudinal hidden photons.

Let us first summarize the details of the search.
The power of the emitter cavity was 47.9~W, and the detector cavity was sensitive to a signal power of $9.8\times 10^{-25}$~W.
The frequency of the mode was 1.74~GHz.
For the axion search a static magnetic field was maintained in both cavities;
this is not necessary for the hidden-photon search and it does not affect the results.
The cavities used in the CROWS experiment were not cylindrical, however to simplify the calculation we approximate them as cylinders of length 6~cm (and radius chosen to give the correct frequency).
Further, the separation between the two cavities was not very large: we take it to be three times the cavity length, but we still use the far-field approximation.
The quality factors of the emitter and detector cavities were 11392 and 12151 respectively; we just take the lower value.
This allows us to calculated the signal strength using Eq.~\ref{eq:receiver-cavity-signal}.
We then take the limit to be set at $|B_{rec}|^2/|B_{em}|^2 = P_{rec}/P_{em} = 2\times10^{-26}$.
We emphasize that these approximations introduce only an order one error, to be compared to the orders-of-magnitude improvement of the limits at low mass. (We also do not recalculate the limit for masses greater than $\omega$, since this will be almost identical to the previous result. A precise calculation could be done using the method presented in section~\ref{sec:cavities}, but is not warranted here.)

The new limit is shown by the solid blue line in Fig.~\ref{fig:reach}. Although the bound is weak compared to preexisting limits, it illustrates the improved low-mass scaling, which will allow large new regions of parameter space to be explored with future experiments.

\subsection{Sensitivity projection for future experiments}
\label{sec:future-sensitivity}

The longitudinal modes produced in the emitter cavity drive modes in a receiver cavity that is resonantly locked to the emitter. The cavity modes excited in the receiver can be read through precision devices such as SQUID magnetometers. The fundamental sources of noise that will limit the sensitivity of this experiment is set by thermal noise in the receiver cavity and the intrinsic noise of the precision device used to read the signal (such as the sensitivity of the SQUID magnetometer). For operating temperatures $T \sim 4$ K, we expect thermal noise to dominate over the intrinsic sensitivity of the magnetometers.

To estimate the sensitivity, the setup can be viewed as an LC receiver circuit that is resonantly driven by a source. The source produces a voltage $V_s$ that is produced at a well defined frequency and phase for an integration time $t_{int}$. Since the LC oscillator has a finite $Q$, it has a non-zero resistance $R$. Thermal (Johnson) noise creates voltage noise across this resistor with a power spectral density equal to $R T$ and a flat dependence on frequency ~\cite{Bradley:2003kg}. Over an integration time $t_{int}$, the Fourier amplitude of this voltage noise in the frequency of interest is $\sqrt{R T (1/t_{int})}$. The signal to noise ratio is thus equal to $\frac{V^2_s t_{int}}{R T}$. Expressing $R$ in terms of $Q$, we see that the signal to noise ratio can be expressed as $\left(P_{signal} \,  t_{int}/ T\right)$ where $P_{signal}$ is the power stored in the receiver.

In the setup discussed in this paper,  $P_{signal} \simeq \frac{\omega B_{rec}^2 V_{cav}}{Q} $, where $V_{cav}$ is the cavity volume.  The signal to noise ratio scales as
\begin{equation}
\label{eqn: snr}
{\rm SNR} \simeq  \frac{P_{signal} \, t_{int}}{T} \simeq  \frac{\omega B_{rec}^2 V_{cav}}{Q}  \frac{t_{int}}{T} \, .
\end{equation}

Taking $B_{rec}$ from
Eq.~\ref{eq:receiver-cavity-signal}, and requiring SNR~$\simgt 5$
gives an estimated reach in the $m_{\gamma'} \ll \omega$ regime of
\begin{equation}
\label{eq:reach}
\varepsilon \simeq 10^{-7.5} \times \left( \frac{10^{-10}~{\rm
eV}}{m_{\gamma'}} \right) \sqrt{\frac{1\,{\rm T}}{B_0}} \left(
\frac{10^{10}}{Q} \right)^\frac{1}{4} \left( \frac{T}{4\,{\rm K}}
\right)^\frac{1}{4} \left( \frac{f}{2.3\rm{GHz}} \right)^\frac{3}{4}
\left( \frac{300 \, {\rm cm}^3}{V_{cav}} \right)^\frac{1}{4} \left(
\frac{1 \, {\rm year}}{t_{int}} \right)^\frac{1}{4} \, ,
\end{equation}
where we have taken the ratio of cavity separation to height to be $d/L=3$.
The red dashed line Fig.~\ref{fig:reach} shows the potential reach of a future cavity experiment with the experimental parameters indicated in Eq.~\ref{eq:reach} \footnote{Thanks to Sami Tantawi for providing realistic cavity parameters.}.

The scaling with integration time in Eq.~\ref{eqn: snr} is more favorable than it would be if we were searching for an unknown signal (as from a background dark matter field for example).  This is because we have control of the emitter and thus know the precise mode driven in the emitter and in particular we know the phase and frequency.  Hence we know the predicted phase and frequency of the signal in the receiver.  Equivalently, we will measure the field amplitude (not power) in the receiver cavity.  This field amplitude builds up linearly in time because of the resonance until a time $\sim Q/f$.  Conceptually, each of these times can be imagined as a separate measurement of the field amplitude.  Then the sensitivity to the field will increase as the square root of the number of such measurements, thus giving the scaling of the sensitivity to be $\sqrt{t_{int}}$ in the field, or $t_{int}$ in the field squared, as shown in Eq.~\ref{eqn: snr}.  With high $Q$ ($\sim 10^{10}$), this scaling with integration time is not a particularly large gain because $Q/f \sim 10$~s, so not very many measurements are being made.  However, for smaller $Q$ ($\sim 10^5$), such as may occur in initial versions of this experiment, this may give a significant improvement in the sensitivity.

\subsection{ALPS}
\label{sec:ALPS}

Of course, the cavities used do not need to be microwave cavities.
It is interesting to consider the use of optical cavities, as in the original Light-Shining-Through-A-Wall setup used in experiments such as ALPS~\cite{Ehret:2009sq, Bahre:2013ywa}.
In principle the frequency of the cavities does not affect the mathematics of our conclusions given above:
the longitudinal mode could still be used as outlined above and would give an enhanced scaling at low masses.

Unfortunately we find that in practice optical cavities do not lend themselves naturally to the type of geometries we have proposed.
A prototypical Light-Shining-Through-A-Wall experiment such as ALPS has two long, resonant, optical cavities, positioned end-to-end but separated by a wall.  One of the cavities is driven by a laser.  There are
several reasons such an experiment does not benefit from the longitudinal mode of the hidden photon.

First, the driving laser itself is not like a normal cavity.  It works by quantum stimulated emission which means it dominantly produces the modes that already have the largest occupation number.  These are, by far, the transverse modes of the photon/hidden photon system, since the normal photon by itself has no longitudinal mode.  Thus the laser will not efficiently produce the longitudinal mode.

Of course the optical cavity it is driving can produce the longitudinal mode, even though it is driven only by transverse modes, exactly like the microwave cavities considered above.  At some level it will produce the longitudinal mode since the mirrors on either end of the optical cavity will have current running in them parallel to the E-field of the driving mode.  This will produce longitudinal radiation with momentum in this same direction (which is perpendicular to the long axis of the cavity), as shown in Figure \ref{fig:edm-radiation-pattern}.  But both the receiver and emitter cavities are long and thin with only small mirrors at each end and lie along one line.  As we see, the longitudinal radiation will be dominantly directed perpendicular to this line and so will couple poorly to the receiver cavity.  This is somewhat similar to the situation of the badly-positioned receiver cavity in Figure \ref{fig:cavity-positioning}.

To improve the transmission of the longitudinal mode one could
locate the receiver cavity next to the emitter, parallel to it but
not along a single line.  This gives a good alignment to the mirrors
which are the walls of the optical cavity, allowing them to
efficiently transmit the longitudinal mode.  However this approach
would lose peak sensitivity because previously the sensitivity was
linearly enhanced by the length of the cavities (the number, $N$, of
wavelengths contained), while in this new geometry that enhancement
would be completely lost (those factors of $N$ are set to 1).  This
loss in peak sensitivity is so great that the enhanced sensitivity
at low masses from the use of the longitudinal mode would not be
sufficient to make this experiment worth doing, especially because,
as can be seen on the sensitivity plot Figure \ref{fig:reach},
extending the reach of ALPS at low mass starting from a reduced peak
sensitivity would only extend the reach in parameter space that is
already covered by other limits.  This remains true even if such
optical Light-Shining-Through-A-Wall experiments are improved by a
few orders of magnitude in the future.

\section{Discussion and conclusions}
\label{sec:conclusions}

A hidden photon coupled to the Standard Model through a kinetic-mixing term is a well motivated possibility for new physics, and a potential dark matter candidate.
We have argued that Light-Shining-Through-A-Wall experiments are a powerful probe of the hidden photon, with a far greater reach than has previously been realized.
In particular, previous work in this direction only made use of the transverse modes of the hidden photon.  In this
paper we have shown that the longitudinal mode allows a large,
parametric improvement in the sensitivity of these experiments.  As
the mass of the hidden photon is decreased below the frequency of
the emitter and receiver, the sensitivity using only the transverse
modes scales as $\varepsilon^2 m_{\gamma'}^4$, but the sensitivity of an
experiment that can make use of the longitudinal mode scales as
$\varepsilon^2 m_{\gamma'}^2$. This is a significant improvement over a huge range of parameter space.

We found that a microwave cavity experiment, for example as shown in Figure \ref{fig:setup}, can take advantage of the sensitivity enhancement from the longitudinal mode.  In order to transmit the longitudinal mode most efficiently between the two cavities, the experiment must be set up with the correct geometry and the correct choice of cavity mode.  The longitudinal mode is dominantly radiated along the direction of the oscillating electric field, perpendicular to the normal transverse modes, as shown in Figure \ref{fig:edm-radiation-pattern}.  Thus, the two cavities must be displaced from each other in a direction parallel to the electric field in the driven mode, as shown in Figure \ref{fig:cavity-positioning}.  With this geometry, the parametric enhancement is attained and the sensitivity can be greatly improved as shown in Figure \ref{fig:reach}.

In fact, this enhancement improves the results of experiments that have already taken place.  For example, the CROWS experiment carried out both a hidden-photon and an axion search.  The hidden-photon search was carried out with a geometry that is sub-optimal for the use of the longitudinal mode.  However, their axion search happened to use a geometry which would efficiently transmit the longitudinal mode of the hidden photon.  Thus the CROWS axion search can in fact be used to place stronger limits on hidden photons than the hidden-photon search.  We have estimated this improved limit and the result is shown in Figure \ref{fig:reach}.

Optical cavity experiments, such as ALPS, could in principle benefit from the longitudinal mode in the same way.  However, in practice such experiments would not gain a useful enhancement because the geometry of these setups does not lend itself to the use of the longitudinal mode (see section \ref{sec:ALPS}).

We estimated the reach of a microwave cavity experiment designed
with optimal geometry to take advantage of the longitudinal mode and high-$Q$ resonant cavities.
Such an experiment could detect hidden photons over many orders of magnitude
of unexplored parameter space in both mass (from $\sim 10^{-18}$ eV to $10^{-4}$ eV) and coupling, as shown in Fig.~\ref{fig:reach}.
This also covers a significant part of the parameter space where the hidden photon can
be the dark matter~\cite{Arias:2012az, future:HPDM}.  This
demonstrates that resonant microwave cavities allow highly sensitive
Light-Shining-Through-A-Wall types of experiments, taking full
advantage of the enhancement from the longitudinal mode of the
hidden photon.  Such an experiment is a powerful probe of the
existence of hidden sectors and new forces in nature.

\section*{Acknowledgements}
We would like to thank Vinod Bharadwaj, Saptarshi Chaudhuri, Kent Irwin, Jeff Neilson, Maxim Pospelov, Josef Pradler, and Sami Tantawi for many useful discussions.  PWG acknowledges the support of NSF grant PHY-1316706, the DOE Early Career Award, the Hellman Faculty Scholars program, and the Terman Fellowship.
JM is supported by a Simons Postdoctoral Fellowship. SR acknowledges the support of ERC grant BSMOXFORD no.~228169 and NSF grant PHY-1417295. YZ is  supported by ERC grant BSMOXFORD no.~228169.

\appendix

\section*{Appendix}

\subsection{Careful derivation of cavity response}
\label{sec:detailed-cavity-derivation}

A chosen cavity mode, of frequency $\omega$, is driven inside the emitter cavity. At $\mathcal O(\varepsilon^0)$, the only fields that exist are those of this driven mode,
\begin{equation}
\vec E(\vec r, t)= \vec E_{em}(\vec r) e^{i \omega t} \, , \qquad \vec B(\vec r, t) = \vec B_{em}(\vec r) e^{i \omega t} \qquad\qquad \text{(fields at }\mathcal O(\varepsilon^0)) \, .
\end{equation}
At $\mathcal O(\varepsilon)$, hidden-photons fields are radiated from the cavity. To find these, it is useful to rewrite the field equations for the photon and hidden-photon fields in a new form. Combining Eqs.~\ref{eq:A-eom} and~\ref{eq:A'-eom} to eliminating $\varrho_{EM}$ and $\vec \jmath_{EM}$, and applying appropriate derivatives, gives
\begin{equation}
(\nabla^2 - \partial_t^2 - m_{\gamma'}^2) (\vec E' - \varepsilon \vec E) = \varepsilon m_{\gamma'}^2 \vec E \, .
\label{eq:E'-eom}
\end{equation}
This is a massive wave equation for the combination $\vec E' - \varepsilon \vec E$ sourced by the $E$-field inside the emitter cavity.
It is easily solved using the Green's function, giving the hidden-photon fields outside the emitter cavity as
\begin{equation}
\vec E'(\vec r, t) = - \varepsilon m_{\gamma'}^2 \int_{\mathcal V_{em}} \!\!d^3 x \, \frac{\vec E_{em}(\vec x)}{4\pi \left| \vec r - \vec x \right|} \, e^{i(\omega t- k \left| \vec r - \vec x \right|)} \qquad\qquad (\vec E' \text{ sourced by } \vec E \text{ at }\mathcal O(\varepsilon)) \, .
\label{eq:E'-radiated-from-cavity}
\end{equation}
Here $\mathcal V_{em}$ is the emitter-cavity volume, and $k^2 \equiv \omega^2 - m_{\gamma'}^2$. For $m_{\gamma'}>\omega$, we make the replacement $k \to -i \kappa$, where $\kappa^2 \equiv m_{\gamma'}^2 - \omega^2$. (Note that $\vec B'(\vec r, t)$ can be found from $\vec B' = (i/\omega) \vec \nabla \times \vec E'$, which is an identity following from the definitions of $\vec E'$ and $\vec B'$.)

The hidden-photon fields penetrate the receiver cavity, where they generate a resonant response of the matching receiver cavity mode at $\mathcal O(\varepsilon^2)$.
This response is determined by Maxwell's equations for the $E$- and $B$-fields inside the receiver cavity, along with the boundary condition Eq.~\ref{eq:conductor-BC}.
To solve for the response, we first write $\vec \nabla \times (\vec \nabla \times \vec E') + \partial_t^2 \vec E' = \vec \nabla (\vec \nabla \cdot \vec E') + (\partial_t^2 - \nabla^2) \vec E =  \vec \nabla (\vec \nabla \cdot \vec E') - m_{\gamma'}^2 \vec E'$, where the second equality follows from Eqs.~\ref{eq:A'-eom} (in vacuum). For the electric field, we have the usual vacuum equation $\vec \nabla \times (\vec \nabla \times \vec E) + \partial_t^2 \vec E = 0$.
Combining these gives a sourced wave equation for $\vec E + \varepsilon \vec E'$ inside the vacuum of the receiver cavity, along with the boundary condition on its interior surface:
\begin{align}
\vec \nabla \times (\vec \nabla \times \big[\vec E + \varepsilon \vec E'\big]) + \partial_t^2 \big[\vec E + \varepsilon \vec E' \big] &=  -\varepsilon \big(m_{\gamma'}^2 \vec E' - \vec \nabla (\vec \nabla \cdot \vec E')\big)  \qquad (\vec E + \varepsilon \vec E' \text{ sourced by }\vec E' ) \\
\qquad\qquad\qquad\qquad\qquad
\big[\vec E + \varepsilon \vec E' \big]_\parallel &= 0
\qquad\qquad\qquad\qquad\qquad\, \text{(B.C. at conducting surface)} \, .
\end{align}

Compare this to the equations governing the electric field sourced by an oscillating current distribution $\vec \jmath\,(\vec r) e^{i \omega t}$ inside a cavity, in standard electromagnetism without a hidden photon:
\begin{align}
\vec \nabla \times (\vec \nabla \times \vec E) + \partial_t^2 \vec E
&= - i \omega \vec \jmath\,(\vec r) e^{i \omega t}
\qquad\qquad\!\!\! (\vec E  \text{ sourced by oscillating current density})
\label{eq:standard-cavity-EoM} \\
\qquad\qquad\qquad\qquad\qquad\qquad \vec E_\parallel &= 0
\qquad\qquad\qquad\qquad\qquad\qquad\qquad \text{(B.C. at conducting surface)} \, .
\label{eq:standard-cavity-BC}
\end{align}
We see that $\vec E + \varepsilon \vec E'$ obeys identical governing equations to $\vec E$ in Eqs.~\ref{eq:standard-cavity-EoM}, \ref{eq:standard-cavity-BC}, with an effective current density given by
\begin{equation}
\vec \jmath_{\rm eff}(\vec r) e^{i \omega t} = - \frac{i \varepsilon}{\omega} \big(m_{\gamma'}^2 \vec E' - \vec \nabla (\vec \nabla \cdot \vec E')\big) \, .
\label{eq:effective-current}
\end{equation}
(Note that Eq.~\ref{eq:standard-cavity-EoM} does not assume $\vec \nabla \cdot \vec E = 0$, being used to describe cavities filled with dielectric materials. It is therefore appropriate to our situation, where $\vec \nabla \cdot [\vec E + \varepsilon \vec E'] \neq 0$.)
Eqs.~\ref{eq:standard-cavity-EoM}, \ref{eq:standard-cavity-BC} have a standard solution found by decomposing $\vec E$ into a complete orthonormal set of basis functions (see, for example, chapter 1 of~\cite{Hill}).
The basis can be chosen as the combination of the (divergence-free) vacuum cavity-mode $E$-fields $\vec E_n(\vec r)$ and a set of irrotational (and divergence-full) basis functions $\vec F_p(\vec r)$,
\begin{equation}
\vec E(\vec r, t) = \sum_n c_n(t) \vec E_n(\vec r) + \sum_p d_p(t) \vec F_p(\vec r)  \, .
\label{eq:mode-decomposition}
\end{equation}
Here $\vec E_n$ and $\vec F_p$ satisfy
\begin{gather}
\vec \nabla \cdot \vec E_n = 0  \qquad \qquad  \nabla^2 E_n = - \omega_n^2 \vec E_n \\
\vec \nabla \times \vec F_p = 0  \qquad \qquad  \nabla^2 F_n = - \tilde \omega_p^2 \vec F_p \\
\int d^3 x \, \vec E_n^* \cdot \vec E_m \propto \delta _{n m}  \qquad \qquad
\int d^3 x \, \vec F_p^* \cdot \vec F_q \propto \delta _{p q}  \qquad \qquad
\int d^3 x \, \vec E_n^* \cdot \vec F_p = 0 \\
\vec E_n \big|_\parallel = \vec F_p \big|_\parallel = 0  \qquad \text{on boundary} \, ,
\label{eq:orthogonality}
\end{gather}
where the integrals in the third line are over the cavity volume, and $\omega_n$ are the frequencies of the cavity modes.

Applying Eqs.~\ref{eq:mode-decomposition}-\ref{eq:orthogonality} to Eq.~\ref{eq:standard-cavity-EoM} gives the equations governing the coefficients in Eq.~\ref{eq:mode-decomposition},
\begin{align}
\ddot c_n(t) + \frac{\omega_n}{Q_n} \dot c_n(t) + \omega_n^2 c_n(t) &= -i \omega \Bigg[ \frac{ \int d^3 x \, \vec E_n^*(\vec x) \cdot \vec \jmath\, (\vec x) }{ \int d^3 x \, |E_n(\vec x)|^2 } \Bigg] e^{i \omega t}
\label{eq:coefficient-ODE-1}\\
\ddot d_p(t) &= -i \omega \Bigg[ \frac{ \int d^3 x \, \vec F_p^*(\vec x) \cdot \vec \jmath\, (\vec x) }{ \int d^3 x \, |F_p (\vec x)|^2 } \Bigg] e^{i \omega t} \, .
\end{align}
Here we have added a damping term by hand in Eq.~\ref{eq:coefficient-ODE-1} to account for the finite $Q$-factors of the cavity modes.
We see that a typical mode is driven at order $c_n \vec E_n \sim d_p \vec F_p \sim \vec \jmath\, / \omega^2$.
However, Eq.~\ref{eq:coefficient-ODE-1} has resonantly enhanced solutions when the driving frequency $\omega$ is close to the frequency of a particular cavity mode.
Writing $\omega = \omega_n + \delta\omega$, this mode's coefficient is given by
\begin{equation}
c_n(t) \xrightarrow{\omega \to \omega_n} - \frac{Q}{\omega + 2 i \, \delta\omega \, Q } \Big(1-e^{-i \delta \omega t -\frac{\omega t}{2 Q}}\Big) \Bigg[ \frac{ \int d^3 x \, \vec E_n^*(\vec x) \cdot \vec \jmath\, (\vec x) }{ \int d^3 x \, |E_n(\vec x)|^2 } \Bigg] e^{i \omega t} \, .
\end{equation}

We will assume that in a cavity-to-cavity hidden-photon search, a single mode in the receiver cavity is tuned to be on resonance with the driven emitter cavity. We denote this mode by $\vec E_{cav}$.
Dropping the non-resonant components,
we therefore find the signal field inside the receiver cavity to be
\begin{equation}
[\vec E + \varepsilon \vec E'](\vec r, t) = - \frac{Q}{\omega + 2 i \, \delta\omega \, Q } \Big(1-e^{-i \delta \omega t -\frac{\omega t}{2 Q}}\Big) \Bigg[ \frac{ \int d^3 x \, \vec E_{cav}^*(\vec x) \cdot \vec \jmath_{\rm eff} (\vec x) }{ \int d^3 x \, |E_{cav}(\vec x)|^2 } \Bigg] \vec E_*(\vec r) e^{i \omega t}  \, .
\label{eq:signal-with-full-t-and-omega-dependence}
\end{equation}

The signal builds up fully after $~2\pi Q$ cycles, i.e. when $t\sim Q/\omega$. In the on-resonance, long time limit ($\delta\omega \ll \omega / Q$ and $t \gg Q/\omega$), the observable signal fields inside the receiver cavity become
\begin{align}
\vec E_{observed}(\vec r, t) \equiv [\vec E + \varepsilon \vec E'](\vec r, t) &= - \frac{Q}{\omega} \Bigg[ \frac{ \int d^3 x \, \vec E_{cav}^*(\vec x) \cdot \vec \jmath_{\rm eff} (\vec x) }{ \int d^3 x \, |E_{cav}(\vec x)|^2 } \Bigg] \vec E_{cav}(\vec r) e^{i \omega t} \\
\vec B_{observed}(\vec r, t) \equiv [\vec B + \varepsilon \vec B'](\vec r, t) &= - \frac{Q}{\omega} \Bigg[ \frac{ \int d^3 x \, \vec E_{cav}^*(\vec x) \cdot \vec \jmath_{\rm eff} (\vec x) }{ \int d^3 x \, |E_{cav}(\vec x)|^2 } \Bigg] \vec B_{cav}(\vec r) e^{i \omega t}  \, ,
\end{align}
where the 2nd line uses $\dot{\vec B}^(\vphantom{E}'^) = - \vec \nabla \times \vec E^(\vphantom{E}'^)$, and $\vec B_{cav} = i \vec \nabla \times \vec E_{cav}/\omega$ is the $B$-field profile of the excited cavity mode.

\subsection{Comparison with previous literature}
\label{sec:previous-literature}

The observation that Light-Shining-Through-A-Wall experiments, originally developed for axion searches, are also sensitive to hidden-photons was first made out in Ref.~\cite{Ahlers:2007rd}. This idea was then developed further in Ref.~\cite{Jaeckel:2007ch} with the proposal to perform such an experiment with resonant microwave cavities, opening up a larger, lower mass region of hidden-photon parameter space. While the fundamental concept of these papers is absolutely correct, both missed the presence of the longitudinal hidden-photon mode and its importance to the power of the experiments.
Therefore, for completeness, in this subsection we briefly compare our calculation to the ones laid out in these papers, pointing out the errors that lead to their incorrect formulae for the signal size.

Ref.~\cite{Ahlers:2007rd} treats the signal in Light-Shining-Through-A-Wall hidden-photon searches as arising purely due to oscillation between the photon and hidden-photon modes (in analogy with the photon-to-axion, axion-to-photon conversion originally searched for by the same experiments).
This approach implicitly includes only the transverse modes of the hidden photon (which is in fact reasonable given the experimental geometry of the laser-based experiments existing then, see section \ref{sec:ALPS}).

Ref.~\cite{Jaeckel:2007ch} begins with an estimation of the signal size based on the results of Ref.~\cite{Ahlers:2007rd}, which therefore does not include the contribution from the longitudinal mode. However, it then follows with a more detailed calculation of the signal based on the field equations, which should in principle have captured the effect of the longitudinal mode.
To describe the subtle error made which prevented this, we here switch to the notation of Ref.~\cite{Jaeckel:2007ch}, which uses the interaction basis for the fields, and uses $\chi$ for the kinetic mixing parameter, $A_\mu$ for the interacting 4-vector potential, and $B_\mu$ for the non-interacting 4-vector potential.

Eq.~19 of Ref.~\cite{Jaeckel:2007ch} uses the Green's function method to give $B_\mu$, sourced by the $A_\mu$ \emph{inside the emitter cavity only}. However, unlike the physical fields $\vec E$ and $\vec B$, the $A_\mu$ field of a driven cavity is also non-zero \emph{outside} the cavity (at least in any standard gauge). $B_\mu$ should therefore be given by Eq.~19, but with the integral now over \emph{all} space. This unbounded integration region must introduce a factor $1/m_{\gamma'}^2$, increasing the scaling of the signal from $m_{\gamma'}^4$ to $m_{\gamma'}^2$. (This is practically difficult to work with, and is avoided by working directly with the physical fields.) In fact, without this change Eq.~19 of Ref.~\cite{Jaeckel:2007ch} is not gauge invariant (note that gauge invariance in the interaction basis involve a transformation of both fields: $A_\mu \to A_\mu + \partial_\mu \Lambda$, $B_\mu \to B_\mu + \chi \partial_\mu \Lambda$).

Ref.~\cite{Jaeckel:2007ch} continues by calculating the response of the receiver cavity to hidden-photon field in a manner essential identical to the one we have used here, but working with the 4-vector fields $B_\mu$, $A_\mu$, rather than the physical fields $\vec E^(\vphantom{E}'^)$, $\vec B^(\vphantom{E}'^)$. The two approaches should be mathematically identical, but the latter has the advantage of making manifest the difference in scaling between transverse and longitudinal modes, as shown here in Eq.~\ref{eq:jeff_trans_vs_long}.

\subsection{Skin depth}
\label{sec:skin-depth}

In this section, we present a calculation of plane wave propagation in
a metal.  We will see that for transverse modes the metal will
dissipate the interaction component of the wave rapidly. The sterile
component is barely affected since its interaction with metal,
compared to the interaction component, is suppressed by
 $\varepsilon$ as well as the small hidden photon mass. Further, the
effect on a longitudinal mode of the hidden photon is negligible.

Let us first consider the transverse wave.  We assume the wave
propagates along the $z$ axis, and we take the polarization of the
electric field to be along the $x$ axis.  Thus by taking Lorentz gauge, we have
\begin{equation} \begin{aligned}
{A_\mu}^{\!\!(}\vphantom{A}'^) =& \, \big( 0, {A_x}^{\!\!(}\vphantom{A}'^) (t,z),0,0 \big) \\
{E_x}^{\!\!(}\vphantom{A}'^) =& -i \omega {A_x}^{\!\!(}\vphantom{A}'^)
\end{aligned}
\qquad\qquad \text{(transverse mode)}
\label{Eq:TransMode}
\end{equation}

With the modified Lorentz force law Eq.~\ref{eq:Lorentz-force}, Ohm's law inside a metal of conductivity $\sigma$ becomes
\begin{equation}
\vec \jmath = \sigma ( \vec E + \varepsilon \vec E') \, .
\end{equation}
Combining this with Eqs.~\ref{eq:A-eom},~\ref{eq:A'-eom}, the equations of motion for ${A_x}^{\!\!(}\vphantom{A}'^)$ inside
the metal can be written as
\begin{equation} \begin{aligned}
(\partial_z^2+\omega^2)A_x =&i \omega \sigma (A_x+\varepsilon A_x') \\
(\partial_z^2+\omega^2-m^2)A_x' =&i \varepsilon \omega \sigma (A_x+\varepsilon A_x') \, .
\end{aligned}
\label{Eq:TransModeEOM}
\end{equation}
There are two eigenmodes of this coupled system,
\begin{equation} \begin{aligned}
A_1 &=& A_x+\frac{\sigma \omega}{\sigma\omega + i m^2}\varepsilon A_x' \simeq A_x+\varepsilon A_x' \\
A_2 &=& A'_x-\frac{\sigma \omega}{\sigma\omega + i m^2}\varepsilon A_x\simeq A'_x-\varepsilon A_x \, ,
\end{aligned}
\label{Eq:TransModeEigen}
\end{equation}
with eigenvalues $(\omega^2-i \sigma \omega)$ and
$(\omega^2-m^2-i \sigma\omega\frac{\varepsilon^2 m^4}{\sigma^2\omega^2+m^4})$ respectively.
Here we only keep the leading term in the $\varepsilon$ expansion.
One can clearly see that the propagation of $A_1$ shares the same
form as normal EM wave in the metal.  Thus when $\sigma\gg\omega\gg
m$, this mode has a skin depth $\delta = 2/\sqrt{\sigma \omega}$.
If the metal is much thicker than the skin depth of this mode, the boundary condition for $A_1$ at its surface is $A_1=0$,
implying $E_{x}+\varepsilon E'_{x}=0$.
On the other hand, the imaginary part of the second eigenvalue is suppressed by a factor of $\varepsilon^2$ as well as
small hidden photon mass. Thus $A_2$ is barely affected by the
metal.

This result is consistent with intuition.  If the incoming planewave
is dominated by $A_x'$, then $A_x$ in the metal is $O(\varepsilon)$.
According to the second line in Eq. (\ref{Eq:TransModeEOM}), one
expects that any perturbations on $A'$ should only show up at
$O(\varepsilon^2)$.

Now let us focus on the propagation of the longitudinal plane wave.
Similar to the transverse case, applying Lorentz gauge, one gets
\begin{equation} \begin{aligned}
{A_\mu}^{\!\!(}\vphantom{A}'^) =& \, \big( {A_t}^{\!\!(}\vphantom{A}'^) (t,z), 0, 0, {A_z}^{\!\!(}\vphantom{A}'^) (t,z) \big) \\
E_z^{(')} =& \frac{1}{i\omega}(\partial_z^2+\omega^2)
{A_z}^{\!\!(}\vphantom{A}'^)
\end{aligned}
\qquad\qquad \text{(longitudinal mode)}
\label{Eq:LongMode}
\end{equation}

The equations of motion for $A_z^{(')}$ inside the metal can be written as
\begin{equation} \begin{aligned}
(\partial_z^2+\omega^2)A_z &=&i \frac{\sigma}{\omega} (\partial_z^2+\omega^2)(A_z+\varepsilon A_z') \\
(\partial_z^2+\omega^2-m^2)A_z' &=&i \varepsilon  \frac{\sigma}{\omega}
(\partial_z^2+\omega^2)(A_z+\varepsilon A_z') \label{Eq:LongModeEOM}
\end{aligned}
\end{equation}
The first line of Eq. (\ref{Eq:LongModeEOM}) can be rewritten as
\begin{equation}
(\partial_z^2+\omega^2)(A_z+\varepsilon A_z')
= \frac{i\omega}{\sigma+i \omega}\varepsilon (\partial_z^2+\omega^2)A_z'
\end{equation}
Substitute into the second line of Eq. (\ref{Eq:LongModeEOM}), one
gets
\begin{eqnarray}
(\partial_z^2+\omega^2-\frac{m^2}{1+\varepsilon^2\sigma/(\sigma+i\omega)})
A_z' =0
\end{eqnarray}
In the limit of $\sigma\gg\omega\gg m$, the eigenvalue can be
written as $(\omega^2-m^2-i \varepsilon^2 \frac{m^2}{\sigma^2}
\sigma\omega )$.  Similar to the transverse mode, the skin depth for
the hidden photon field is much larger than normal EM skin depth
because of small $\varepsilon^2$ and small hidden photon mass.

\bibliography{arxiv_v1.bib}

\providecommand{\href}[2]{#2}\begingroup\raggedright\begin{thebibliography}{10}

\bibitem{Holdom:1986eq}
B.~Holdom, ``{Searching for Epsilon Charges and a New U(1)},''
\href{http://dx.doi.org/10.1016/0370-2693(86)90470-3}{{\em Phys.Lett.}
  {\bfseries B178} (1986) 65}.

\bibitem{Pospelov:2007mp}
M.~Pospelov, A.~Ritz, and M.~B. Voloshin, ``{Secluded WIMP Dark Matter},''
  \href{http://dx.doi.org/10.1016/j.physletb.2008.02.052}{{\em Phys.Lett.}
  {\bfseries B662} (2008) 53--61},
\href{http://arxiv.org/abs/0711.4866}{{\ttfamily arXiv:0711.4866 [hep-ph]}}.

\bibitem{Abel:2008ai}
S.~Abel, M.~Goodsell, J.~Jaeckel, V.~Khoze, and A.~Ringwald, ``{Kinetic Mixing
  of the Photon with Hidden U(1)S in String Phenomenology},''
  \href{http://dx.doi.org/10.1088/1126-6708/2008/07/124}{{\em JHEP} {\bfseries
  0807} (2008) 124},
\href{http://arxiv.org/abs/0803.1449}{{\ttfamily arXiv:0803.1449 [hep-ph]}}.

\bibitem{ArkaniHamed:2008qn}
N.~Arkani-Hamed, D.~P. Finkbeiner, T.~R. Slatyer, and N.~Weiner, ``{A Theory of
  Dark Matter},'' \href{http://dx.doi.org/10.1103/PhysRevD.79.015014}{{\em
  Phys.Rev.} {\bfseries D79} (2009) 015014},
\href{http://arxiv.org/abs/0810.0713}{{\ttfamily arXiv:0810.0713 [hep-ph]}}.

\bibitem{ArkaniHamed:2008qp}
N.~Arkani-Hamed and N.~Weiner, ``{Lhc Signals for a Superunified Theory of Dark
  Matter},'' \href{http://dx.doi.org/10.1088/1126-6708/2008/12/104}{{\em JHEP}
  {\bfseries 0812} (2008) 104},
\href{http://arxiv.org/abs/0810.0714}{{\ttfamily arXiv:0810.0714 [hep-ph]}}.

\bibitem{Pospelov:2008zw}
M.~Pospelov, ``{Secluded U(1) Below the Weak Scale},''
  \href{http://dx.doi.org/10.1103/PhysRevD.80.095002}{{\em Phys.Rev.}
  {\bfseries D80} (2009) 095002},
\href{http://arxiv.org/abs/0811.1030}{{\ttfamily arXiv:0811.1030 [hep-ph]}}.

\bibitem{Goodsell:2009xc}
M.~Goodsell, J.~Jaeckel, J.~Redondo, and A.~Ringwald, ``{Naturally Light Hidden
  Photons in Large Volume String Compactifications},''
  \href{http://dx.doi.org/10.1088/1126-6708/2009/11/027}{{\em JHEP} {\bfseries
  0911} (2009) 027},
\href{http://arxiv.org/abs/0909.0515}{{\ttfamily arXiv:0909.0515 [hep-ph]}}.

\bibitem{Arvanitaki:2009hb}
A.~Arvanitaki, N.~Craig, S.~Dimopoulos, S.~Dubovsky, and J.~March-Russell,
  ``{String Photini at the Lhc},''
  \href{http://dx.doi.org/10.1103/PhysRevD.81.075018}{{\em Phys.Rev.}
  {\bfseries D81} (2010) 075018},
\href{http://arxiv.org/abs/0909.5440}{{\ttfamily arXiv:0909.5440 [hep-ph]}}.

\bibitem{Jaeckel:2010ni}
J.~Jaeckel and A.~Ringwald, ``{The Low-Energy Frontier of Particle Physics},''
  \href{http://dx.doi.org/10.1146/annurev.nucl.012809.104433}{{\em
  Ann.Rev.Nucl.Part.Sci.} {\bfseries 60} (2010) 405--437},
\href{http://arxiv.org/abs/1002.0329}{{\ttfamily arXiv:1002.0329 [hep-ph]}}.

\bibitem{Essig:2010ye}
R.~Essig, J.~Kaplan, P.~Schuster, and N.~Toro, ``{On the Origin of Light Dark
  Matter Species},'' {\em Submitted to Physical Review D} (2010) ,
\href{http://arxiv.org/abs/1004.0691}{{\ttfamily arXiv:1004.0691 [hep-ph]}}.

\bibitem{Ringwald:2012hr}
A.~Ringwald, ``{Exploring the Role of Axions and Other Wisps in the Dark
  Universe},'' \href{http://dx.doi.org/10.1016/j.dark.2012.10.008}{{\em
  Phys.Dark Univ.} {\bfseries 1} (2012) 116--135},
\href{http://arxiv.org/abs/1210.5081}{{\ttfamily arXiv:1210.5081 [hep-ph]}}.

\bibitem{Nelson:2011sf}
A.~E. Nelson and J.~Scholtz, ``{Dark Light, Dark Matter and the Misalignment
  Mechanism},'' \href{http://dx.doi.org/10.1103/PhysRevD.84.103501}{{\em
  Phys.Rev.} {\bfseries D84} (2011) 103501},
\href{http://arxiv.org/abs/1105.2812}{{\ttfamily arXiv:1105.2812 [hep-ph]}}.

\bibitem{Arias:2012az}
P.~Arias, D.~Cadamuro, M.~Goodsell, J.~Jaeckel, J.~Redondo, {\em et~al.},
  ``{Wispy Cold Dark Matter},''
  \href{http://dx.doi.org/10.1088/1475-7516/2012/06/013}{{\em JCAP} {\bfseries
  1206} (2012) 013},
\href{http://arxiv.org/abs/1201.5902}{{\ttfamily arXiv:1201.5902 [hep-ph]}}.

\bibitem{future:HPDM}
P.~W. Graham, J.~Mardon, S.~Rajendran, and Y.~Zhao {\em to appear} .

\bibitem{Ade:2014xna}
{\bfseries BICEP2 Collaboration} Collaboration, P.~Ade {\em et~al.}, ``{Bicep2
  I: Detection of B-Mode Polarization at Degree Angular Scales},''
\href{http://arxiv.org/abs/1403.3985}{{\ttfamily arXiv:1403.3985
  [astro-ph.CO]}}.

\bibitem{Holdom:1985ag}
B.~Holdom, ``{Two U(1)'s and Epsilon Charge Shifts},''
\href{http://dx.doi.org/10.1016/0370-2693(86)91377-8}{{\em Phys.Lett.}
  {\bfseries B166} (1986) 196}.

\bibitem{Reece:2009un}
M.~Reece and L.-T. Wang, ``{Searching for the Light Dark Gauge Boson in
  Gev-Scale Experiments},''
  \href{http://dx.doi.org/10.1088/1126-6708/2009/07/051}{{\em JHEP} {\bfseries
  0907} (2009) 051},
\href{http://arxiv.org/abs/0904.1743}{{\ttfamily arXiv:0904.1743 [hep-ph]}}.

\bibitem{Batell:2009di}
B.~Batell, M.~Pospelov, and A.~Ritz, ``{Exploring Portals to a Hidden Sector
  Through Fixed Targets},''
  \href{http://dx.doi.org/10.1103/PhysRevD.80.095024}{{\em Phys.Rev.}
  {\bfseries D80} (2009) 095024},
\href{http://arxiv.org/abs/0906.5614}{{\ttfamily arXiv:0906.5614 [hep-ph]}}.

\bibitem{Bjorken:2009mm}
J.~D. Bjorken, R.~Essig, P.~Schuster, and N.~Toro, ``{New Fixed-Target
  Experiments to Search for Dark Gauge Forces},''
  \href{http://dx.doi.org/10.1103/PhysRevD.80.075018}{{\em Phys.Rev.}
  {\bfseries D80} (2009) 075018},
\href{http://arxiv.org/abs/0906.0580}{{\ttfamily arXiv:0906.0580 [hep-ph]}}.

\bibitem{Aubert:2009af}
{\bfseries BaBar Collaboration} Collaboration, B.~Aubert {\em et~al.},
  ``{Search for a Narrow Resonance in E+E- to Four Lepton Final States},''
\href{http://arxiv.org/abs/0908.2821}{{\ttfamily arXiv:0908.2821 [hep-ex]}}.

\bibitem{deNiverville:2011it}
P.~deNiverville, M.~Pospelov, and A.~Ritz, ``{Observing a Light Dark Matter
  Beam with Neutrino Experiments},''
  \href{http://dx.doi.org/10.1103/PhysRevD.84.075020}{{\em Phys.Rev.}
  {\bfseries D84} (2011) 075020},
\href{http://arxiv.org/abs/1107.4580}{{\ttfamily arXiv:1107.4580 [hep-ph]}}.

\bibitem{Hewett:2012ns}
J.~Hewett, H.~Weerts, R.~Brock, J.~Butler, B.~Casey, {\em et~al.},
  ``{Fundamental Physics at the Intensity Frontier},''
\href{http://arxiv.org/abs/1205.2671}{{\ttfamily arXiv:1205.2671 [hep-ex]}}.

\bibitem{Dharmapalan:2012xp}
{\bfseries MiniBooNE Collaboration} Collaboration, R.~Dharmapalan {\em et~al.},
  ``{Low Mass WIMP Searches with a Neutrino Experiment: a Proposal for Further
  Miniboone Running},''
\href{http://arxiv.org/abs/1211.2258}{{\ttfamily arXiv:1211.2258 [hep-ex]}}.

\bibitem{Ahlers:2007rd}
M.~Ahlers, H.~Gies, J.~Jaeckel, J.~Redondo, and A.~Ringwald, ``{Light from the
  Hidden Sector},'' \href{http://dx.doi.org/10.1103/PhysRevD.76.115005}{{\em
  Phys.Rev.} {\bfseries D76} (2007) 115005},
\href{http://arxiv.org/abs/0706.2836}{{\ttfamily arXiv:0706.2836 [hep-ph]}}.

\bibitem{Jaeckel:2007ch}
J.~Jaeckel and A.~Ringwald, ``{A Cavity Experiment to Search for Hidden Sector
  Photons},'' \href{http://dx.doi.org/10.1016/j.physletb.2007.11.071}{{\em
  Phys.Lett.} {\bfseries B659} (2008) 509--514},
\href{http://arxiv.org/abs/0707.2063}{{\ttfamily arXiv:0707.2063 [hep-ph]}}.

\bibitem{Caspers:2009cj}
F.~Caspers, J.~Jaeckel, and A.~Ringwald, ``{Feasibility, Engineering Aspects
  and Physics Reach of Microwave Cavity Experiments Searching for Hidden
  Photons and Axions},''
  \href{http://dx.doi.org/10.1088/1748-0221/4/11/P11013}{{\em JINST} {\bfseries
  4} (2009) P11013},
\href{http://arxiv.org/abs/0908.0759}{{\ttfamily arXiv:0908.0759 [hep-ex]}}.

\bibitem{Povey:2010hs}
R.~Povey, J.~Hartnett, and M.~Tobar, ``{Microwave Cavity Light Shining Through
  a Wall Optimization and Experiment},''
  \href{http://dx.doi.org/10.1103/PhysRevD.82.052003}{{\em Phys.Rev.}
  {\bfseries D82} (2010) 052003},
\href{http://arxiv.org/abs/1003.0964}{{\ttfamily arXiv:1003.0964 [hep-ex]}}.

\bibitem{Wagner:2010mi}
A.~Wagner, G.~Rybka, M.~Hotz, L.~Rosenberg, S.~Asztalos, {\em et~al.}, ``{A
  Search for Hidden Sector Photons with ADMX},''
  \href{http://dx.doi.org/10.1103/PhysRevLett.105.171801}{{\em Phys.Rev.Lett.}
  {\bfseries 105} (2010) 171801},
\href{http://arxiv.org/abs/1007.3766}{{\ttfamily arXiv:1007.3766 [hep-ex]}}.

\bibitem{Betz:2013dza}
M.~Betz, F.~Caspers, M.~Gasior, M.~Thumm, and S.~Rieger, ``{First Results of
  the Cern Resonant Wisp Search (Crows)},''
  \href{http://dx.doi.org/10.1103/PhysRevD.88.075014}{{\em Phys.Rev.}
  {\bfseries D88} (2013) 075014},
\href{http://arxiv.org/abs/1310.8098}{{\ttfamily arXiv:1310.8098
  [physics.ins-det]}}.

\bibitem{Ehret:2010mh}
K.~Ehret, M.~Frede, S.~Ghazaryan, M.~Hildebrandt, E.-A. Knabbe, {\em et~al.},
  ``{New Alps Results on Hidden-Sector Lightweights},''
  \href{http://dx.doi.org/10.1016/j.physletb.2010.04.066}{{\em Phys.Lett.}
  {\bfseries B689} (2010) 149--155},
\href{http://arxiv.org/abs/1004.1313}{{\ttfamily arXiv:1004.1313 [hep-ex]}}.

\bibitem{Arias:2010bh}
P.~Arias, J.~Jaeckel, J.~Redondo, and A.~Ringwald, ``{Optimizing
  Light-Shining-Through-A-Wall Experiments for Axion and Other Wisp
  Searches},'' \href{http://dx.doi.org/10.1103/PhysRevD.82.115018}{{\em
  Phys.Rev.} {\bfseries D82} (2010) 115018},
\href{http://arxiv.org/abs/1009.4875}{{\ttfamily arXiv:1009.4875 [hep-ph]}}.

\bibitem{Horns:2012jf}
D.~Horns, J.~Jaeckel, A.~Lindner, A.~Lobanov, J.~Redondo, {\em et~al.},
  ``{Searching for Wispy Cold Dark Matter with a Dish Antenna},''
  \href{http://dx.doi.org/10.1088/1475-7516/2013/04/016}{{\em JCAP} {\bfseries
  1304} (2013) 016},
\href{http://arxiv.org/abs/1212.2970}{{\ttfamily arXiv:1212.2970}}.

\bibitem{Seviour:2014dqa}
R.~Seviour, I.~Bailey, N.~Woollett, and P.~Williams, ``{Hidden-Sector Photon
  and Axion Searches Using Photonic Band Gap Structures},''
\href{http://dx.doi.org/10.1088/0954-3899/41/3/035005}{{\em J.Phys.} {\bfseries
  G41} (2014) 035005}.

\bibitem{An:2013yfc}
H.~An, M.~Pospelov, and J.~Pradler, ``{New Stellar Constraints on Dark
  Photons},''
\href{http://arxiv.org/abs/1302.3884}{{\ttfamily arXiv:1302.3884 [hep-ph]}}.

\bibitem{An:2013yua}
H.~An, M.~Pospelov, and J.~Pradler, ``{Dark Matter Detectors as Dark Photon
  Helioscopes},''
\href{http://arxiv.org/abs/1304.3461}{{\ttfamily arXiv:1304.3461 [hep-ph]}}.

\bibitem{Redondo:2013lna}
J.~Redondo and G.~Raffelt, ``{Solar Constraints on Hidden Photons
  Re-Visited},'' \href{http://dx.doi.org/10.1088/1475-7516/2013/08/034}{{\em
  JCAP} {\bfseries 1308} (2013) 034},
\href{http://arxiv.org/abs/1305.2920}{{\ttfamily arXiv:1305.2920 [hep-ph]}}.

\bibitem{Redondo:2008aa}
J.~Redondo, ``{Helioscope Bounds on Hidden Sector Photons},''
  \href{http://dx.doi.org/10.1088/1475-7516/2008/07/008}{{\em JCAP} {\bfseries
  0807} (2008) 008},
\href{http://arxiv.org/abs/0801.1527}{{\ttfamily arXiv:0801.1527 [hep-ph]}}.

\bibitem{Ehret:2009sq}
{\bfseries ALPS collaboration} Collaboration, K.~Ehret {\em et~al.},
  ``{Resonant Laser Power Build-Up in Alps: a `Light-Shining-Through-Walls'
  Experiment},'' \href{http://dx.doi.org/10.1016/j.nima.2009.10.102}{{\em
  Nucl.Instrum.Meth.} {\bfseries A612} (2009) 83--96},
\href{http://arxiv.org/abs/0905.4159}{{\ttfamily arXiv:0905.4159
  [physics.ins-det]}}.

\bibitem{Bahre:2013ywa}
R.~Bähre, B.~Döbrich, J.~Dreyling-Eschweiler, S.~Ghazaryan, R.~Hodajerdi,
  {\em et~al.}, ``{Any Light Particle Search II —Technical Design Report},''
  \href{http://dx.doi.org/10.1088/1748-0221/8/09/T09001}{{\em JINST} {\bfseries
  8} (2013) T09001},
\href{http://arxiv.org/abs/1302.5647}{{\ttfamily arXiv:1302.5647
  [physics.ins-det]}}.

\bibitem{Sami}
S.~Tantawi {\em private communication} .

\bibitem{Bradley:2003kg}
R.~Bradley, J.~Clarke, D.~Kinion, L.~Rosenberg, K.~van Bibber, {\em et~al.},
  ``{Microwave Cavity Searches for Dark-Matter Axions},''
\href{http://dx.doi.org/10.1103/RevModPhys.75.777}{{\em Rev.Mod.Phys.}
  {\bfseries 75} (2003) 777--817}.

\bibitem{Hill}
D.~A. Hill, {\em {Electromagnetic Fields in Cavities}}.
\newblock Wiley, 2009.

\end{thebibliography}\endgroup
\bibliographystyle{utphys}

\end{document}